\documentclass[twocolumn,fleqn,showpacs,showkeys,superscriptaddress]{revtex4}
\usepackage[pdftex]{graphicx}
\usepackage{amssymb}
\usepackage{amsmath}
\usepackage{bm}
\usepackage{kotex}
\usepackage{multirow}
\usepackage{array}
\usepackage{url}
\usepackage{hyperref}

\newcommand\arcdeg{\mbox{$^\circ$}}%
\newcommand\degr{\arcdeg}%
\let\la=\lesssim        

\newcommand\ion[2]{#1$\;${%
\ifx\@currsize\normalsize\small \else
\ifx\@currsize\small\footnotesize \else
\ifx\@currsize\footnotesize\scriptsize \else
\ifx\@currsize\scriptsize\tiny \else
\ifx\@currsize\large\normalsize \else
\ifx\@currsize\Large\large
\fi\fi\fi\fi\fi\fi
\rmfamily\@Roman{#2}}\relax}%

\begin{document}
\setcounter{page}{0}
\title[]{Space Missions for Astronomy and Astrophysics in Korea: Past, Present, and Future}
\thanks{Authors contributing to individual missions are co-first authors.}

\author{Kwang-il \surname{Seon}}
\email{kiseon@kasi.re.kr}
\thanks{contributed to FIMS/SPEAR}
\affiliation{Korea Astronomy and Space Science Institute, Daejeon 34055, Republic of Korea}

\author{Wonyong \surname{Han}}
\email{whan@kasi.re.kr}
\thanks{contributed to MIRIS}
\affiliation{Korea Astronomy and Space Science Institute, Daejeon 34055, Republic of Korea}

\author{Young-Wook \surname{Lee}}
\email{ywlee2@yonsei.ac.kr}
\thanks{contributed to GALEX}
\affiliation{Department of Astronomy and Center for Galaxy Evolution Research, Yonsei University, Seoul 03722, Republic of Korea}

\author{Hyung Mok \surname{Lee}}
\email{hmlee@kasi.re.kr}
\thanks{contributed to AKARI}
\affiliation{Korea Astronomy and Space Science Institute, Daejeon 34055, Republic of Korea}

\author{Min Bin \surname{Kim}}
\email{happykmb@gmail.com}
\thanks{contributed to UFFO}
\affiliation{Department of Physics, Sungkyunkwan University, Suwon 16419, Republic of Korea}

\author{I. H. \surname{Park}}
\affiliation{Department of Physics, Sungkyunkwan University, Suwon 16419, Republic of Korea}

\author{Woong-Seob \surname{Jeong}}
\email{jeongws@kasi.re.kr}
\thanks{contributed to NISS and SPHEREx}
\affiliation{Korea Astronomy and Space Science Institute, Daejeon 34055, Republic of Korea}

\author{Kyung-Suk \surname{Cho}}
\email{kscho@kasi.re.kr}
\thanks{contributed to NGC}
\affiliation{Korea Astronomy and Space Science Institute, Daejeon 34055, Republic of Korea}

\author{Jae Jin \surname{Lee}}
\email{jjlee@kasi.re.kr}
\thanks{contributed to SNIPE}
\affiliation{Korea Astronomy and Space Science Institute, Daejeon 34055, Republic of Korea}

\author{Dae-Hee \surname{Lee}}
\email{dhlee@kasi.re.kr}
\thanks{contributed to UVOMPIS}
\affiliation{Korea Astronomy and Space Science Institute, Daejeon 34055, Republic of Korea}
\affiliation{Department of Aerospace Engineering, Korea Advanced Institute of Science and Technology, Daejeon 34141, Republic of Korea}


\author{Kyujin \surname{Kwak}}
\email{kkwak@unist.ac.kr}
\affiliation{Department of Physics, Ulsan National Institute of Science and Technology, Ulsan 44919, Republic of Korea}


\date[]{}

\begin{abstract}
We review the history of space mission in Korea focusing on the field of astronomy and astrophysics. For each mission, scientific motivation and achievement are reviewed together with some technical details of the program including mission schedule. This review includes the ongoing and currently approved missions as well as some planned ones. Within the admitted limitations of authors' perspectives, some comments on the future direction of space program for astronomy and astrophysics in Korea are made at the end of this review.    
\end{abstract}



\maketitle

\section{INTRODUCTION}

Astronomy is one of the oldest subjects in natural science. In some sense, it is older than physics and chemistry because looking up the (night) sky, observing what is there, and thinking why it is are doing ``science" by definition. Humans have been doing these scientific activities since the very beginning of the human history. Night sky is always up there. Physics joined astronomy soon because it provided answers to why things on sky look like them from the knowledge that it obtained by watching the Mother Nature. For this reason, it could be said that astrophysics is the earliest sub-field of physics. 

In modern astronomy and astrophysics, astronomical are collected with tools. Historically, optical telescopes have been steadily developed since the first observation of Galilei and advanced to the level of state of the art. Currently, three  large optical telescopes with a diameter of about thirty meter diameters are under construction: Thirty Meter Telescope (TMT), Giant Magellan telescope (GMT), and Extremely Large Telescope (ELT). These telescopes are also capable of observing near-infrared sources and equipped with high resolution spectrographs and adaptive optics which can increase the spatial resolution close to the diffraction limit of the telescope by reducing the atmospheric seeing. 
Korea Astronomy and Space Science Institute (KASI) is one of the founding institutes of the GMT project with $10\%$ share. 

Radio telescopes are another important tool in observational astronomy. The development of radio telescope followed the rapid advancement in the radar technology during the World War II. Since then, radio telescopes have advanced further to the current state-of-the-art level. The interferometry technique is optimal for radio telescopes, and it became recently possible for the earth size telescope arrays called ``Event Horizon Telescope (EHT)" to take a direct image of a black hole centered in the nearby galaxy M87\cite{2019ApJ...875L...1E}. Korean radio astronomers also participated in this historic observation 
%
%
 by utilizing the Korean VLBI Network (KVN: operated by KASI), the East Asian VLBI Network (EAVN:  operated by the consortium of China, Japan and Korea), and James Clerk Maxwell Telescope (JCMT: operated by the East Asian Observatory [EAO]). EAO is formed by the East Asian Core Observatories Association (EACOA)  composed of ASIAA (Academia Sinica Institute of Astronomy and Astrophysics) of Taiwan, NAOC (National Observatories of Chinese Academy of Sciences), NAOJ (National Astronomical Observatory of Japan), and KASI. KASI is also an operating partner of the Atacama Large Millimeter Array (ALMA), which is a major telescope facility for the EHT collaboration. Square Kilometer Array (SKA) will be the next major radio interferometer and KASI is currently one of the observers.

Since the earth atmosphere allows electromagnetic wave only in the optical and radio band to reach its surface, space mission is required to obtain data in the other wavelengths. Since (and thanks to) the furious competition for space exploration between the Soviet Union (former Russia) and the USA during 1960s, numerous worldwide space missions for astronomy and astrophysics have been designed, proposed, launched, and accomplished. X-ray astronomy began as a result of these various space programs, especially thanks to the rocket engineering and the X-ray detector technology. The importance of X-ray astronomy as an emerging field of astronomy in addition to traditional fields of optical and radio was recognized with the Nobel prize awarded to Riccardo Giacconi in 2002.

%

It is now undebatable that space missions are an essential part of modern astronomy and astrophysics.  They cover wider ranges of wavelengths than ground-based telescopes and often have better performance  even in the same wavelength covered by the ground-based telescopes. However, the costs of space missions are generally much higher than those of ground-based telescopes and for this reason, only a few countries have been able to afford to operate space programs. So far, the National Aeronautics and Space Administration (NASA) of the USA has been dominating the field of space missions. An exemplary program of NASA is a ``great observatories program" which covers a wide range of wavelengths with four high-cost satellites: the Hubble Space Telescope (HST: near-ultraviolet/optical/near-infrared), Compton Gamma Ray Observatory (CGRO: gamma-ray), the Chandra X-ray Observatory (CXO: X-ray), and the Spitzer Space Telescope (SST: infrared). Among these, HST and CXO are still operating after they were launched in 1990 and 1999, respectively. Except SST, the other three satellites were deployed from the space shuttle. The service missions to repair/upgrade HST were also done with the space shuttle program. NASA also has explorers program which supports mid- and low-cost space missions\footnote{\url{https://explorers.gsfc.nasa.gov}}. The NASA homepage\footnote{\url{https://www.nasa.gov/missions}} lists all its missions including those with the international collaboration. 

European countries sometimes have their own space missions, but have been putting more collaborative efforts through the European Space Agency (ESA). Japan has been also developing several space missions and the Japan Aerospace Exploration Agency (JAXA) governs the space programs in Japan. Thanks to the rich experiences during the space exploration era, Russia and other countries of previous Soviet Union still have strong space programs with relevant technologies. India and China also try to promote their space programs.  
Due to the high-cost of space mission, 
international collaboration has been pursued more seriously nowadays, particularly for large missions that also require expertise. Many missions are now operated and/or planned under the international collaboration (sometimes between agencies).

Space program in Korea has begun in early 1990s by launching the first satellite called ``Wooribyul-1" (our star in Korean, which is also named KITSAT-1) in August 1992. A series of small satellite missions followed. Some of them were dedicated to astronomy and astrophysics. The small satellites, whose missions primarily focused on science and technology, were separately named STSAT. For example, STSAT-1 (also called FIMS/SPEAR: 
Far-ultraviolet Imaging Spectrogram / Spectroscopy of Plasma Evolution from Astrophysical Radiation), 
the first space mission of Korean space program for astronomy and astrophysics, is often called ``Wooribyul-4", the 4th mission in the small satellite series. There are two more (successful) space missions for astronomy and astrophysics in the small satellite series: STSAT-3 (also called MIRIS: Multi-purpose InfraRed Imaging System) and NEXTSAT-1 (also called NISS: Near-infrared Imaging Spectrometer for Star formation history) which is named to indicate the NEXT-generation satellite program, continuation of the small satellite series. Conventionally, satellites are categorized with their weights: small ($< 500$~kg), medium ($500-1000$~kg), and large ($>1000$~kg). 

For about three decades after KITSAT-1, Korea has been developing and operating various satellites, most of which are for broadcasting and communication (KOREASAT: a series of satellites also called ``Mugunghwa", named after Korean national flower) and for multiple purposes (KOMPSAT: a series of satellites also called ``Arirang", named after the most popular Korean traditional song). Korean space program also succeeded in operating geo-stationary satellites: communication, ocean and meteorological satellite (COMS: also called ``Cheollian-1", telegnosis in Korean) and GEO-KOMPSAT (also called ``Cheollian", continuation as a series of geo-stationary satellites). The Korea Aerospace Research Institute (KARI) has been playing the key role of governing the space programs in Korea (i.e., developing and operating most satellites) and the Satellite Technology Research Center (SaTReC) in the Korea Advanced Institute of Science and Technology (KAIST) also contributed, in particular, to the small satellites. Contributions to space missions for astronomy and astrophysics have been made by KASI. 
Korean space program has been also trying to develop its own technologies required for the missions which include development of Korea Space Launch Vehicle (KSLV) and construction of Naro Space Center. 
In addition to aforementioned programs, a new program called CAS500 (500 kg-class Compact Advanced Satellites) is under operation and some of these missions are for astronomy and astrophysics. 

This review first describes early Korean space missions dedicated to astronomy and astrophysics: STSAT-1 (FIMS/SPEAR: Section \ref{FIMS_sec}) and STSAT-3 (MIRIS: Section \ref{MIRIS_sec}). Then we move onto the past international collaboration space missions which Korean astronomers/astrophysicists participated in: Galaxy Evolution Explorer (GALEX: Section \ref{GALEX_sec}), AKARI (Section \ref{AKARI_sec}), and Ultra-Fast Flash Observatory (UFFO: Section \ref{UFFO_sec}). This review also covers the ongoing and planned missions which are dedicated or related to astronomy and astrophysics. Section \ref{NISS_SPHEREx_sec} describes both the past Korean space mission (NEXTSAT-1: NISS) and the ongoing/planned mission in collaboration with NASA (SPHEREx: Spectro-Photometer for the History of the Universe, Epoch of Reionization and Ices Explorer) which was recently selected as a NASA Medium Explorer (MIDEX) mission. Another ongoing/planned mission in collaboration with NASA is presented in Section \ref{NGC_sec}, where CODEX (COronal Diagnostic EXperiment), a next generation coronagraph, which will be installed on the International Space Station (ISS), will be introduced together with the relevant previous experiments. Sections \ref{SNIPE_sec} and \ref{UVOMPIS_sec} describe SNIPE (Small scale magNetospheric and Ionospheric  Plasma Experiment) and UVOMPIS (UV-Optical Multiband Polarizing Imager System), respectively, both of which are part of the CAS500 program although SNIPE is a mission with cube (or micro) satellites. 
%
%
%
%
%
%
%
%
%
%
%
Individual mission sections provide descriptions on (1) background and scientific motivation, (2) history or schedule of mission, (3) achievements, and (4) connections to or perspectives for following/future missions. In the last section (Section \ref{conclusion_sec}), we will make some comments on the future direction of  Korean space missions for astronomy and astrophysics based upon our experiences in the past and ongoing missions.

\section{FIMS/SPEAR: STSAT-1}\label{FIMS_sec}



\begin{figure}
\includegraphics[width=8.5cm]{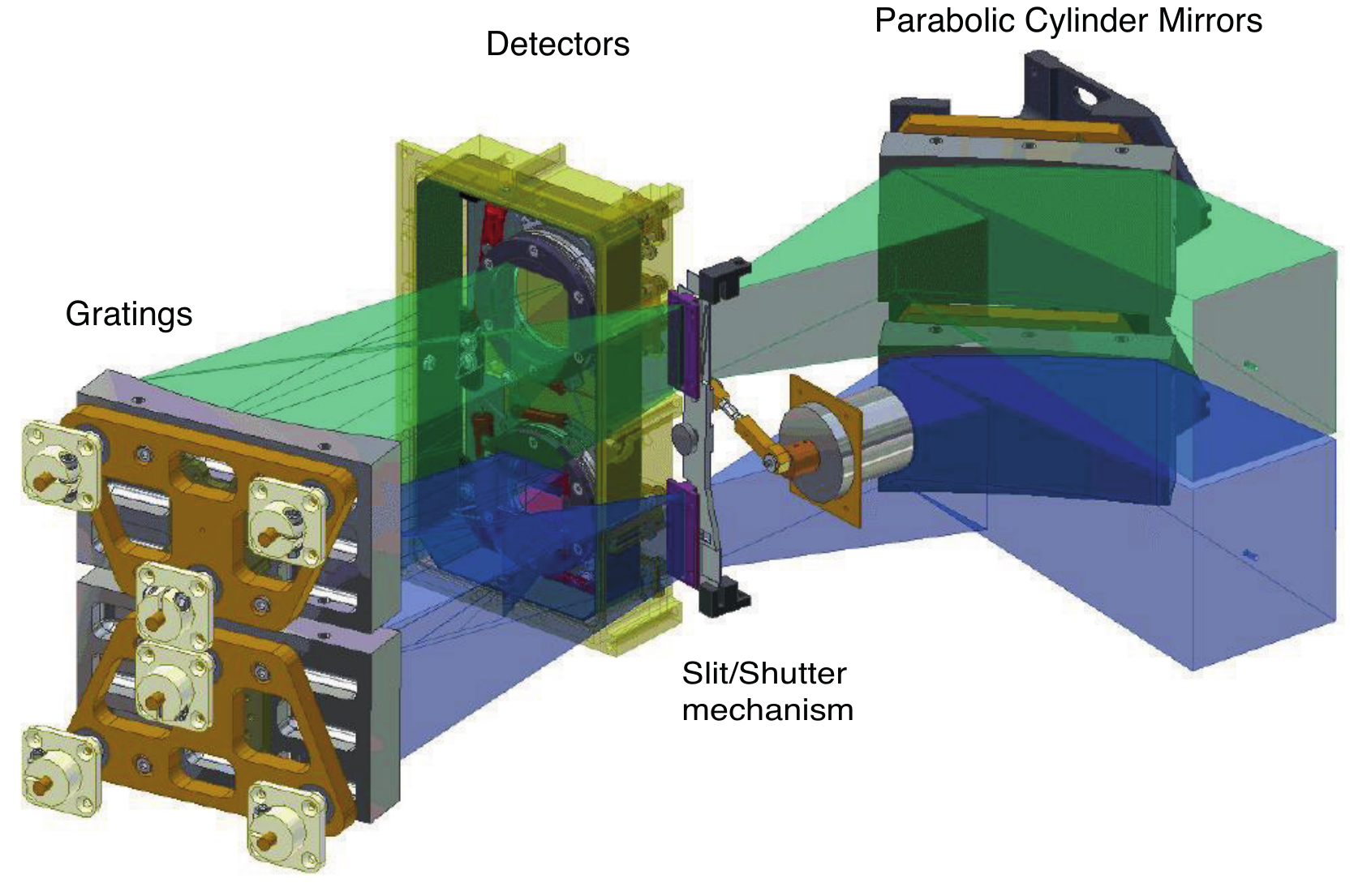}
\caption{Optical Layout of FIMS/SPEAR dual-spectrograph. Plane-parallel light is collected by the mirrors to a slit shutter. The light diffracts from ellipsoidal gratings to the position-sensitive photon counting detectors. Adopted from \cite{2006ApJ...644L.159E}.
}\label{fig_FIMS_optics}
\end{figure}

FIMS/SPEAR is the primary payload on the STSAT-1,  which was the first Korean science satellite, launched on 2003 September 27 by Russian launcher COSMOS \cite{2006ApJ...644L.153E,2006ApJ...644L.159E}.
The STSAT-1 also carried four additional instruments designed to acquire the in-situ measurements of space plasma on the Earth’s polar region (an Electrostatic Analyzer, Solid State Telescopes, a Langmuir Probe, and a Scientific Magnetometer)~\cite{2014JASS...31..311L}. 
The FIMS/SPEAR  mission  was  a  joint  project  of  KASI,  KAIST, and the University of California at Berkeley (UC Berkeley).

The official development period of the STSAT-1, including the FIMS/SPEAR instrument, was from October 1997 to December 2003, five years and four months, and the total research fund was about 11.7 billion Korean won.
The KAIST and UC Berkeley research team explored the possibilities of FIMS/SPEAR and prepared its conceptual design and basic optical design several years before the official start of STSAT-1 project.
As soon as the STSAT-1 project began, the Korean research team, KASI and KAIST, worked with SaTReC's satellite development team to define the mission, operational policy, and design and develop the optical, mechanical, and electronic systems. Overall system design and the development of micro-channel plate (MCP) detectors was led by UC Berkeley in the United States. Detailed implementations of the optomechanical and electronic systems, including design, test, manufacturing, and operation, were led by the Korean team. The off-axis parabolic cylindrical mirror was the part which had been tried first in astronomy, and techniques both for manufacturing and precision measurement had been newly devised in Korea.

The development process was not without difficulties. In particular, in the final thermal vacuum test of the FIMS flight model, one of the two MCP detectors malfunctioned and could not be resolved in Korea, so the detector was sent back to the United States for repair. However, in the process of re-entering Korea, the detector could not be restored to its normal state, so the sensitivity in the short wavelength channel of the system became severely lowered.
The STSAT-1 was launched into a 700~km Sun-synchronous orbit at 98$^\circ$.2 inclination with an orbital period of 98.5 minutes and a $\sim$34 minute eclipse. About 10 daily orbits were scheduled for astronomy observations. One or two orbits per day were used for observations of the northern nightside aurora.
Since November 2004, a problem had occurred in the battery of the satellite system, and the observation had been intermittently attempted, but after May 2005, observations were no longer possible.

The FIMS/SPEAR space mission had conducted the first large-scale spectral survey of the sky in the far ultraviolet (FUV) wavelengths (900-1750~\AA). Its primary scientific objective was to observe strong cooling lines from warm and hot (transition-temperature, $T=10^4-10^6$~K) gas,  fluorescence emission lines from molecular hydrogen (H$_2$), and dust-scattered starlight in the Milky Way galaxy. FIMS/SPEAR also observed the earth airglow and auroral emission lines.

FIMS/SPEAR is composed of dual imaging spectrographs optimized for the measurement of diffuse FUV emission. The two spectral channels are referred to as the ``short'' wavelength band (900--1150~\AA; $4^\circ.0\times4'.6$ field of view) and ``long'' wavelength band (1350--1750~\AA; $7^\circ.4\times4'.3$). Each spectrograph has a spectral resolution of $\lambda/\Delta\lambda\sim550$
and an imaging resolution of $5'-10'$. Each channel is composed of a collecting mirror, a diffraction grating, and a position-sensitive photon-counting MCP detector. The channels share the same slit/shutter and detector focal plane. As shown in Fig. \ref{fig_FIMS_optics}, a cylindrical parabolic mirror focuses incident plane-parallel light from the sky to a slit. The cylindrical wave from the slit then illuminates an ellipse of rotation surface with constant grating rulings, which focuses an imaged spectrum to a MCP detector. The cylindrical-source scheme provides a large grasp (defined as a product of solid angle and collecting area) optimized for diffuse source. The MCPs share a single read-out and electronics system, which utilize a cross delay line anode system. The broad bandpass, spectral resolution, large field of view and imaging resolution were chosen to detect a variety of important astrophysical emission lines from abundant ionic species while rejecting intense airglow line contamination, such as hydrogen Ly$\alpha$ 1216~\AA~and neutral oxygen O I 1304~\AA.

The FIMS/SPEAR mission had observed $\sim80$\% of the whole sky. The all-sky survey observation had been performed by scanning the entire sky along the short axis of the slit, i.e., along the $\sim5'$ field direction of the $\sim8^\circ\times 5'$ FOV (field of view). In each orbit, the FOV scanned $\sim180^\circ$, from the North ecliptic pole to the South ecliptic pole, during the eclipse interval ($\sim25$ minutes) of the satellite orbit, and the pole-to-pole scanning drifts $360^\circ$ along the ecliptic equator for one year because of the properties of a Sun-synchronous orbit.

Thanks to the unique design of FIMS/SPEAR optimized for diffuse sources, a variety of unprecedented fruitful results were obtained. The main achievements are summarized along with their scientific importance.

\subsection{Highly Ionized, Transition-Temperature Gas}

The interstellar medium (ISM) plays a key role in the evolutionary cycle of matter in galaxies. The ISM can be classified into a few, approximately stable phases in pressure equilibrium \cite{1977ApJ...218..148M}: the cold neutral medium (CNM) with a temperature of $\sim10^2$~K, the warm neutral/ionized medium (WNM/WIM) with a temperature of $\sim10^4$~K, and the hot ionized medium (HIM) with a temperature of $\sim10^6$~K. Since Spitzer \cite{1956ApJ...124...20S} predicted the existence of the diffuse hot gas in the Galactic halo, there has been many attempts for the origin and evolution of the hot gas. One of the essential keys in understanding the hot gas evolution is to measure the global spatial distribution and volume-filling fraction of the transition-temperature ($10^4<T<10^6$~K), cooling gas in the Milky Way. The cooling gas could provide direct tool to understand interaction between the HIM and WIM/WNM.

\begin{figure}
\includegraphics[width=8.5cm]{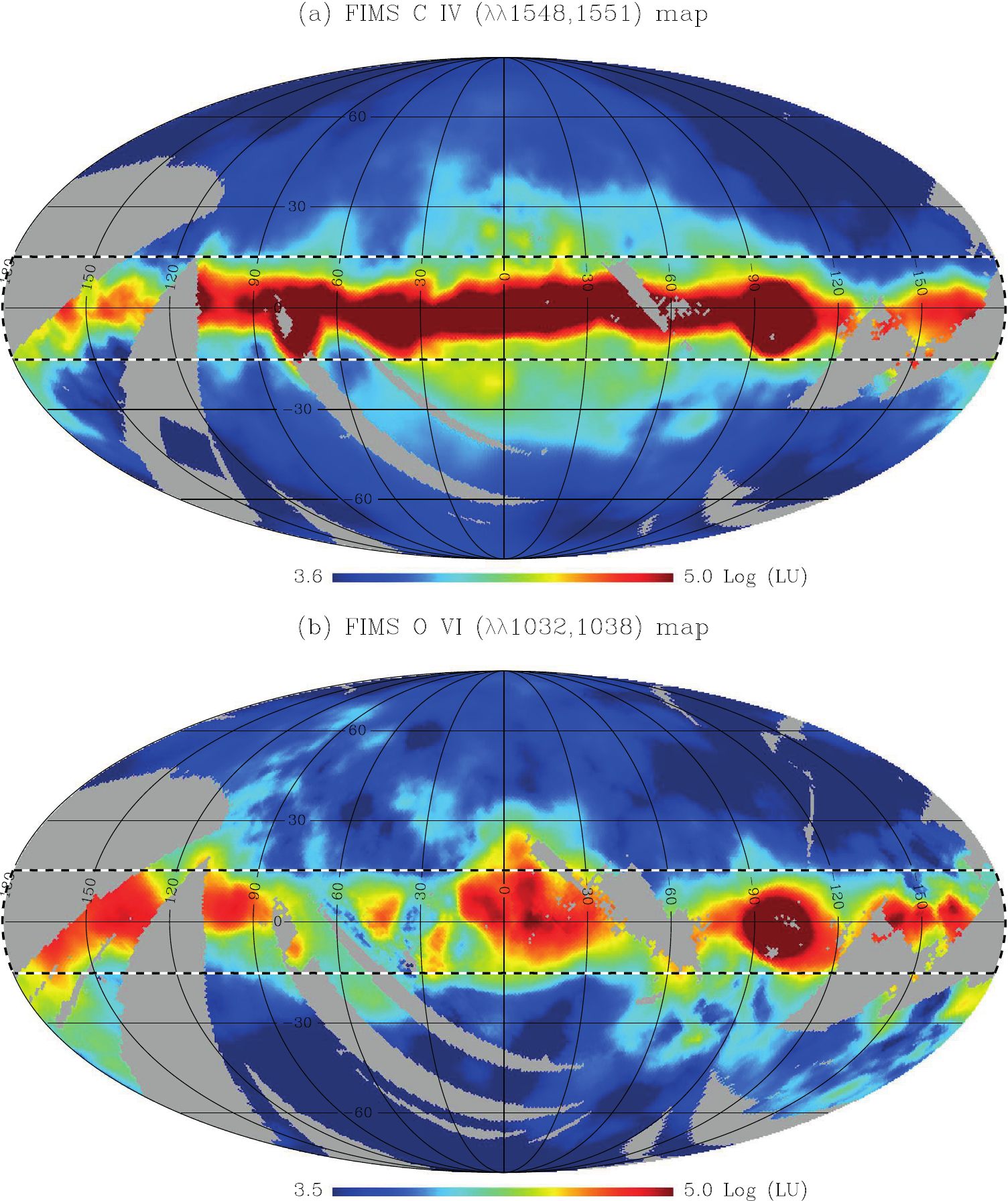}
\caption{(a) C IV all-sky map and (b) O VI map after correcting for dust extinction. The strong extinction region of $\left|b \right|<15^\circ$ around the Galactic disk is indicated with a dashed rectangular box. Here, LU denotes line unit (photons~cm$^{-2}$ s$^{-1}$ sr$^{-1}$). Adopted from \cite{2019ApJS..243....9J}.
}\label{fig_CIV_OVI_map}
\end{figure}

The transition-temperature gas cools down by emitting resonance lines of highly-ionized atoms predominantly in the FUV wavelengths: for instance, O VI $\lambda\lambda1031, 1038$~\AA, Si IV $\lambda\lambda1394, 1403$~\AA, and C IV $\lambda\lambda1548, 1551$~\AA. Observations of these FUV lines have been carried out mainly through the ``absorption'' line studies using FUV space missions. However, the absorption line measurements can provide information only along limited lines of sight toward bright stars or quasars. Therefore, observations of the FUV ``emission'' lines for the entire sky are crucial to understand the global scale properties of the transition-temperature gas.

FIMS/SPEAR obtained, for the first time, the all-sky maps of two main FUV cooling lines, C IV and O VI (Fig. \ref{fig_CIV_OVI_map}) \cite{2019ApJS..243....9J}. The temperature of the cooling gas was found to be more or less uniform throughout the Milky Way with a value of $(1.89\pm0.06)\times10^5$ K. Assuming a simple disk-like model, the scale height of the electron density was found to be $z_{\rm e}=6^{+3}_{-2}$ kpc, and the total mass of the transition-temperature gas was estimated to be $\sim 6.4^{+5.2}_{-2.8}\times 10^9~ M_{sun}$. The volume-filling fraction of the gas was estimated to be $\left<f\right>=0.26\pm0.09$ on average, ranging from $f\sim0.3$ in the inner Galaxy to $f\sim0.18$ in the outer Galaxy. It was also found that the overall distribution of C IV and O VI cannot be explained by traditional models, such as a simple supernova remnant (SNR) model or a three-phase model. Instead, the intensity ratio of C IV to O VI could be explained by combined effects of SNRs and turbulent mixing layers. Thermal conduction front models and high-velocity cloud models also appear to be consistent with the FIMS/SPEAR observation.

\begin{figure}
\includegraphics[width=8.3cm]{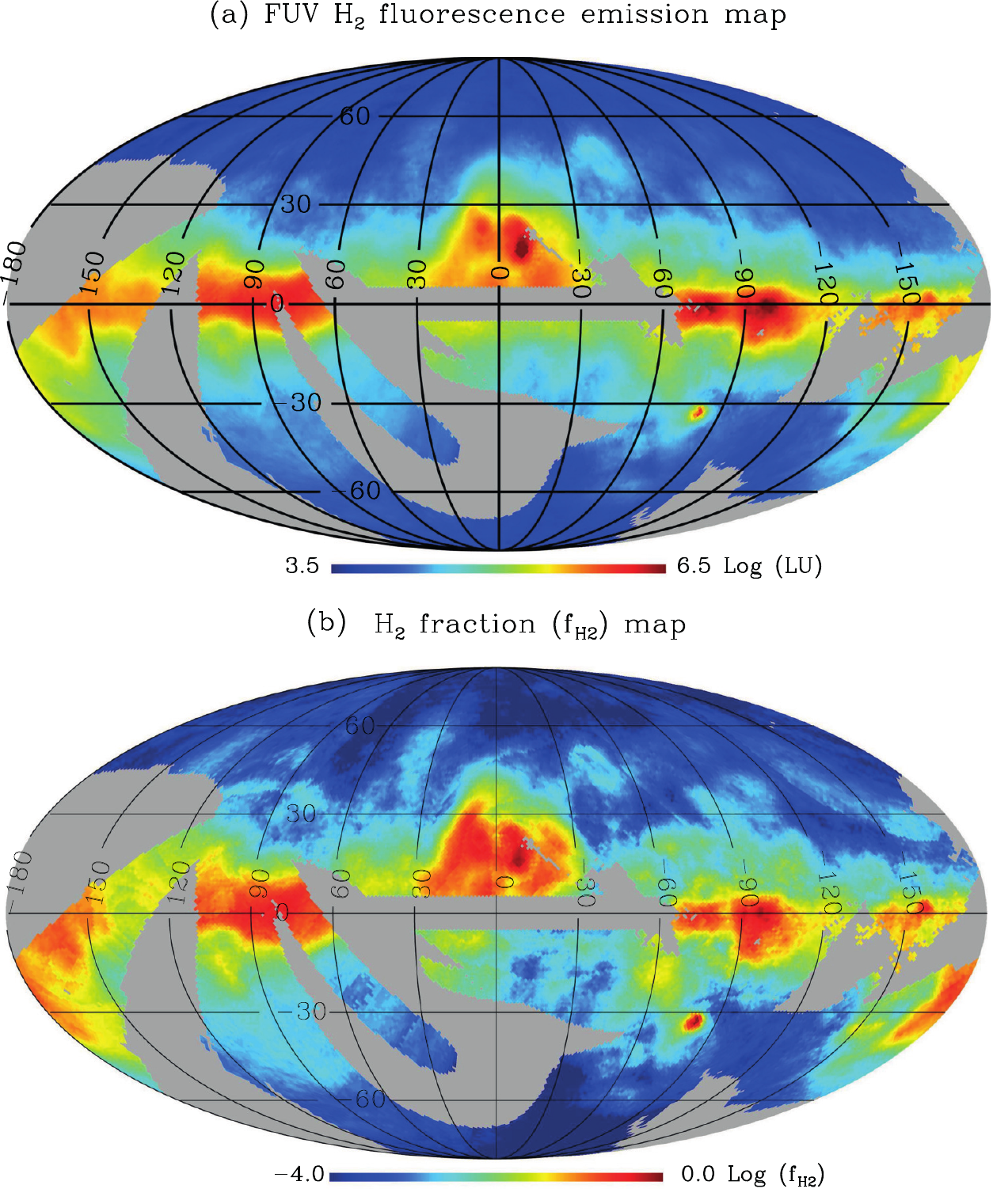}
\caption{(a) FUV fluorescent H$_2$ emission line map after correcting for dust extinction and (b) all-sky map of the H$_2$ fraction $f_{\rm H_2}$. The strongly extinguished region of $\left|\ell\right|<60^\circ$ and $\left|b\right|<5^\circ$ is excluded in the figure. Adopted from \cite{2017ApJS..231...21J}.
}\label{fig_H2_map}
\end{figure}

\subsection{Molecular Hydrogen Gas}

Molecular clouds are the birthplace of stars, and the newly born stars radiate intense ultraviolet (UV) photons which govern the chemical and the physical properties of the adjacent molecular clouds. Hydrogen is the most abundant element in the universe and $\sim20$\% of the interstellar hydrogen is in the form of H$_2$; H$_2$ is therefore an essential tracer for star-forming regions and molecular clouds. However, it is difficult to observe H$_2$ directly from the ground because of the absence of a permanent electric dipole moment \cite{1971ApJ...163..155H,1982ARA&A..20..163S}.

H$_2$ molecules are electronically excited by absorbing FUV photons in the Lyman ($B^1\Sigma_u^+ - X^1\Sigma_g^+$) and Werner ($C^1\Pi^u - X^1\Sigma_g^+$) bands. Subsequent fluorescence leads to dissociation of the H$_2$ molecules in about 10\% of the cases and de-excitation to vibrationally excited states of the ground electronic state in the remaining 90\% of the cases.
When the fluorescent transitions from the electronically excited state to the vibration-rotation states in the electronic ground state occur, a series of FUV emission lines in the Lyman and Werner bands are emitted. The cascade within the vibrational rotational levels of the ground electronic state result in quadrupole transition lines at the near- and mid-infrared wavelengths \cite{1989ApJ...347..863S}. However, there have been no near-infrared or FUV observational studies of fluorescent H$_2$ emission lines over the entire sky. Ground-based telescope observations detected H$_2$ near-infrared fluorescence emission lines only in a small number of sight lines.  Since the FUV wavelengths are only accessible from the space telescope, only a few observations of these H$_2$ emission have been reported before FIMS/SPEAR.  The FIMS/SPEAR mission was the first which provided the opportunity for the all-sky survey of H$_2$ fluorescence lines.

Jo et al. \cite{2017ApJS..231...21J} presented, for the first time, the all-sky map of FUV fluorescent H$_2$ emission in the Milky Way, which covers $\sim76$\% of the sky (Fig. \ref{fig_H2_map}). The H$_2$ fluorescence emission was found to contribute to, on average, $\sim8.7\% \pm 2.4\%$ of the total FUV flux in the wavelength range of $1370-1710$~\AA~ observed by FIMS/SPEAR. They found that, after the dust-extinction correction, the H$_2$ fluorescence lines correlate well with the color excess $E(B-V)$, the atomic hydrogen column density N(HI), and hydrogen H$\alpha$ emission. A simple plane-parallel photodissociation region model was compared with the FIMS/SPEAR data to estimate the spatial distribution of H$_2$ molecules. In the model, a uniform pressure of $p/k=1000$ cm$^{-3}$ K was assumed and the three-dimensional interstellar radiation field (ISRF) was calculated using the stellar models for the bright UV stars. The study revealed that the H$_2$ column density N(H$_2$) ranges between $10^{16}$ and $10^{23}$ cm$^{-2}$. The molecular fraction of hydrogen ($f_{\rm H_2}$) was found to be ranging from $<1\%$ at optically thin regions with $E(B-V)<0.1$ to $\sim50\%$ in the Galactic plane with $E(B-V)=3$; these results indicate strong self-shielding and active formation of H$_2$ in the Galactic plane. They also found the mean gas-to-dust ratio to be $\sim5.1\times10^{21}$ atoms cm$^{-2}$ mag$^{-1}$~\cite{2017ApJS..231...21J}, which is consistent with the standard value of $\sim5.8\times10^{21}$ atoms cm$^{-2}$ mag$^{-1}$, which was derived from the previous FUV space missions \emph{Corpernicus} \cite{1978ApJ...224..132B} and \emph{Far Ultraviolet Spectroscopic Explorer (FUSE)} \cite{2009ApJS..180..125R}, within the uncertainty of the data. However, the ratio appears to be higher than the standard value in the Galactic halo region.

\begin{figure}
\includegraphics[width=8.3cm]{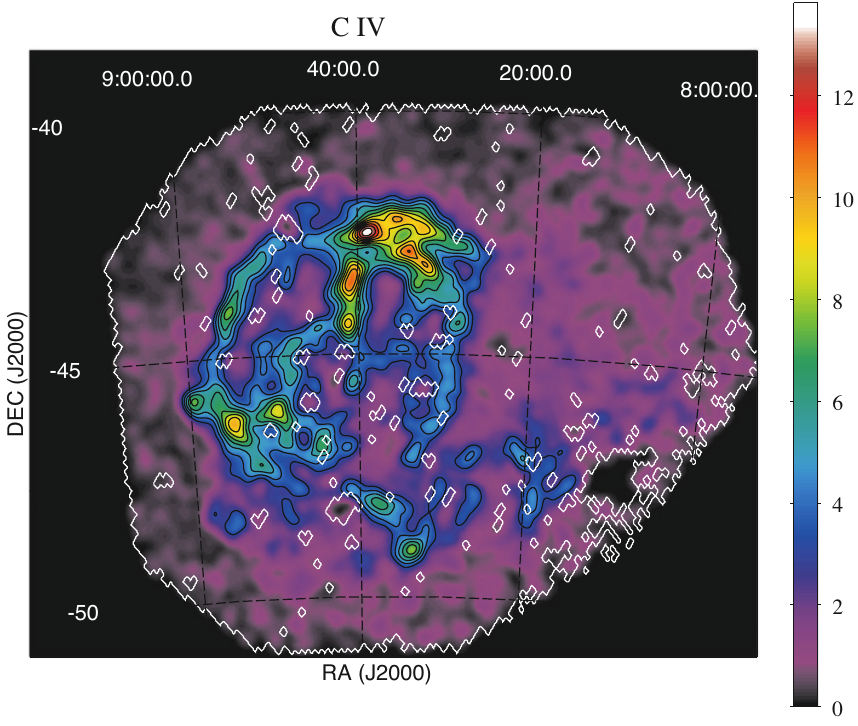}
\caption{C IV $\lambda\lambda~1548, 1551$ emission-line image of the Vela SNR. The units of the color bar is $10^{-6}$ erg s$^{-1}$ cm$^{-2}$ sr$^{-1}$. The white solid lines indicate the pixels that were masked to remove bright stars. The map was not corrected for dust extinction. Adopted from \cite{2012ApJ...761..135K}.
}\label{fig_Vela_SNR}
\end{figure}

\subsection{Diffuse Galactic Light}

Measurement of the ISRF, a sum of the direct starlight and the diffuse galactic light (DGL), is of considerable interest because the ISRF, in particular at the FUV wavelengths, plays a major role in controlling the physics and the chemistry of the ISM. Historically, there had been long debates on the origin of the DGL in the FUV \cite{1991ARA&A..29...59B,1991ARA&A..29...89H}. The DGL is currently believed to be mainly Galactic in origin, starlight scattered off the interstellar dust. Therefore, observations of the DGL are able to give us critical information on the scattering properties and spatial distribution of the interstellar dust.

The FIMS/SPEAR mission obtained the first, large-scale spectroscopic map of the DGL in the FUV \cite{2011ApJS..196...15S,2019ApJS..243....9J}. Seon et al. \cite{2011ApJS..196...15S} extensively analyzed the FUV DGL and found a good correlation between the FUV DGL and other quantities that trace the ISM, such as H I column density, 100$\mu$m far-infrared thermal emission from interstellar dust grains, H$\alpha$ emission from photoionized hydrogen gas. Interestingly, the FUV DGL was found to correlate better with the diffuse H$\alpha$ emission than with other Galactic ISM tracers. The strong linear correlation of the FUV DGL with the diffuse H$\alpha$, especially in low intensities, are likely to indicate that the dust-scattered component of H$\alpha$ becomes increasingly important at high latitudes; this seems to provide an important clue on the nature of the diffuse WIM \cite{2011ApJ...743..188S,2012ApJ...758..109S}. It was also found that the spectrum of the FUV DGL is in general flat, but slightly softer than the direct stellar spectrum. The relative softening of the FUV DGL compared with the direct starlight is likely due to the rise of dust albedo at longer wavelengths.

\subsection{Individual Extended Objects}
 In addition to the above all-sky survey results, FIMS/SPEAR also had conducted deep pointed observations toward selected targets, such as SNRs \cite{2006ApJ...644L.175S,2006ApJ...644L.171N,2006ApJ...644L.189S,2007ApJ...670.1132S,2007ApJ...665L.139K,2010ApJ...709..823K,2010ApJ...722..388K,2012ApJ...761..135K,2014ApJ...784...12K}, superbubbles \cite{2006ApJ...644L.167K,2007ApJ...665L..39P,2008ApJ...678L..29R,2007A&A...472..509W,2012ApJ...756...38J,2015ApJ...807...68J}, H II regions \cite{2010ApJ...719.1964P,2013ApJ...774...34C,2015ApJ...800..132C}, and H$_2$ molecular clouds \cite{2006ApJ...644L.181L,2006ApJ...644L.185R,2008ApJ...686.1155L,2009ApJ...700..155P,2012ApJ...754...10P,2013ApJ...765..107L,2015MNRAS.449..605L}. The observations on the individual objects also have yielded a variety of interesting results, which could be obtained only through a FUV space mission optimized for diffuse sources.
 
 For instance, Kim et al. \cite{2012ApJ...761..135K} constructed the FUV spectral images for the entire Vela SNR (Fig. \ref{fig_Vela_SNR}), at various emission lines including C III $\lambda~977$, O VI $\lambda\lambda~1032, 1038$, Si IV $\lambda\lambda~1394, 1403$, O IV] $\lambda~1404$, C IV $\lambda~1548, 1551$, He II $\lambda~1640.5$, and O III] $\lambda\lambda~1661, 1666$. The results are invaluable in that the previous space missions could observe only small patches of such a huge SNR. They also found an interesting FUV enhancement in an east region of the Vela SNR, which seems to be caused by a contribution from non-radiative shocks of another very young SNR, the Vela Jr. SNR that overlaps with the Vela SNR. This is the first detection of the FUV signal that is likely associated with the Vela Jr. SNR, supporting previous arguments that it is too far from us. Kregenow et al. \cite{2006ApJ...644L.167K} observed the Orion-Eridanus superbubble and found direct observational evidence on an evolved thermal interface between hot gas inside the bubble and the cooler ambient medium, which has been theoretically anticipated. A catalog of the spectra of bright stars observed during the sky survey was also made as a by-product \cite{2016MNRAS.456..417J}.

\section{MIRIS: STSAT-3: Compact Infrared Space Telescope}\label{MIRIS_sec}


\subsection{Background}

The first Korean infrared space telescope MIRIS was developed as the main payload of the Korean Science and Technology Satellite (STSAT-3), by KASI in collaboration with SaTReC of KAIST, KARI, and ISAS (Institute of Space and Astronautical Science of Japan).  Following the successful launch of the STSAT-1, the space science payload RFP (request for proposal) for the STSAT-3 was released by MOST (Ministry of Science and Technology in Korea) in 2007 and the MIRIS project proposal was selected as the main payload of the satellite among several proposals with open competition base.  The MIRIS was developed in 5 years and delivered to KARI in 2013.  After extensive assembly tests, the STSAT-3 was successfully launched by Russian Dnepr rocket in November 2013 into a Sun-synchronous orbit of altitude around 600 km.  The main science mission of the MIRIS was near-infrared wide-field observations for the Galactic plane, North Ecliptic Pole (NEP), and the North Galactic Pole (NGP).  Three narrowband filters (Paschen-$\alpha$ [Pa$\alpha$] line and Pa$\alpha$ dual continuum) were employed, in addition to two broadband filters.  The narrow band filters were used for observations along the Galactic plane with the latitude of $-3\degr \la b \la$$+3\degr$, while the broad band filters were used for investigating the cosmic infrared background radiation.  After initial verification operations in early 2014, MIRIS observed the Galactic plane as scheduled until the expiration of the cryogenic cooler in May 2015, producing the Pa$\alpha$ emission line data of Galactic plane survey, and the $\it{I}$ and  $\it{H}$ wideband data for NEP and NGC to detect the cosmic infrared background (CIB) radiation.

\begin{figure}
\includegraphics[width=8.3cm]{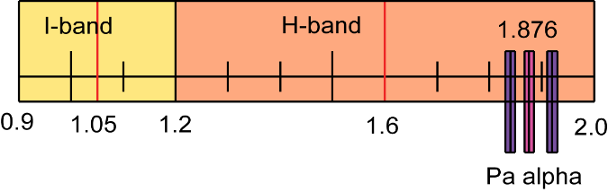}
\caption{The bandpass of the MIRIS filter system for broadband $\it{I}$ and $\it{H}$, and narrowband Pa$\alpha$ with a dualband continuum filter. Adopted from \cite{2014PASP..126..853H}.
}\label{fig_MIRIS_1}
\end{figure}

\begin{table}
\caption{Specifications of MIRIS (adopted from~\cite{2014PASP..126..853H})}
\begin{ruledtabular}
\begin{tabular}{ll}
Parameters & Specifications \\
\colrule
Aperture      & {80~mm}                   \\
F-number             & {F/2}            \\
Effective Focal Length                 & {160~mm}       \\
Lens Material      & {S-FPL53,  S-TIH6, Fused Silica}        \\
Wavelength            & {0.9$\sim$2~$\mu$m}                   \\
Filters                & {Broadband: $\it{I}$ (1.05 $\mu$m), $\it{H}$(1.6 $\mu$m)}  \\
                        & {Narrowband: Pa$\alpha$ (1.876 $\mu$m)}           \\
                        & {Dual-band Continuum (1.84, 1.92 $\mu$m)}           \\
Limiting Magnitude& {$\it{I}$:18.6, $\it{H}$:19.1    }        \\
Sensor               & {Teledyne PICNIC (HgCdTe)}          \\
Sensor Format        & {256$\times$256 pixels}       \\
Pixel Size               & {40~$\mu$m}                            \\
Detector FOV           & {3.67$^\circ$$\times$3.67$^\circ$}       \\
Pixel FOV               & {51.6{$\rm''$}}                          \\
Readout Noise        & {45 e$^{-}$}                          \\
Telescope Temperature & {180$\sim$200~K}  \\
Sensor Temperature      & {90~K}                            \\
Compatible Orbit            & {595 $ \sim$ 635 km}         \\
Total Mass               & {29.8 kg}                            \\
\end{tabular}
\end{ruledtabular}
\label{MIRIS_table1}
\end{table}

\subsection{MIRIS System Design}

The initial MIRIS design considered many constraints, particularly its mass  ($<$30 kg), length ($<$70 cm) and power consumption ($<$30 W) within the satellite platform. Wide-field survey observation with short focus optical design was examined for the science missions. Near-infrared observations  require specific sensor for appropriate response at particular infrared wavelength bands, among those available from the industry, coupled with infrared optical filters.  The telescope system and  sensor needed cryogenic cooling for proper response at infrared wavelength bands.  The opto-mechanical design was critical for proper support of all necessary components, while the electronics designs addressed proper interface with infrared sensor and the spacecraft bus.

A short focus optical design was effective at arranging the wide-field of view for the survey observations.  The telescope system was designed as F/2 refractive optics, and comprised of 5 lenses including a 80~mm diameter main lens.   Coupled with the pixel size of 40~$\mu$m Teledyne 256$\times$256 PICNIC infrared sensor, the design can obtain pixel scale of {51.6{$\rm''$}}, totaling 3.67$^\circ$$\times$3.67$^\circ$ field of view, sufficiently wide for the Galactic plane survey observations in a year.  Though when compared to 2K$\times$2K infrared sensors that recently became available, the Teledyne 256$\times$256 infrared sensor is relatively limited in its size, it was a widely available option from the industry at the time of development.  In total, five infrared filters were employed for astronomical observations. Two broadband filters of $\it{I}$ band (0.8$ \sim$1.35 $\mu$m) and  $\it{H}$ band (1.15$\sim$2.05 $\mu$m) were used for CIB, while two narrowband filters, Pa$\alpha$ (1.876 $\mu$m) and a specially designed dual-band continuum filter (1.84+1.92 $\mu$m) were used for the Galactic plane survey.  One additional blank filter was prepared for calibration.  The filter wheel and driving system were designed to house these filters with one mechanical switch and three magnetic sensors.  Fig. \ref{fig_MIRIS_1} shows the bandpass of the MIRIS filters for broadband $\it{I}$ and  $\it{H}$, narrowband Pa$\alpha$, and  dualband continuum filter \cite{2014PASP..126..853H}. 

\begin{figure*}[t]
\includegraphics[width=14cm]{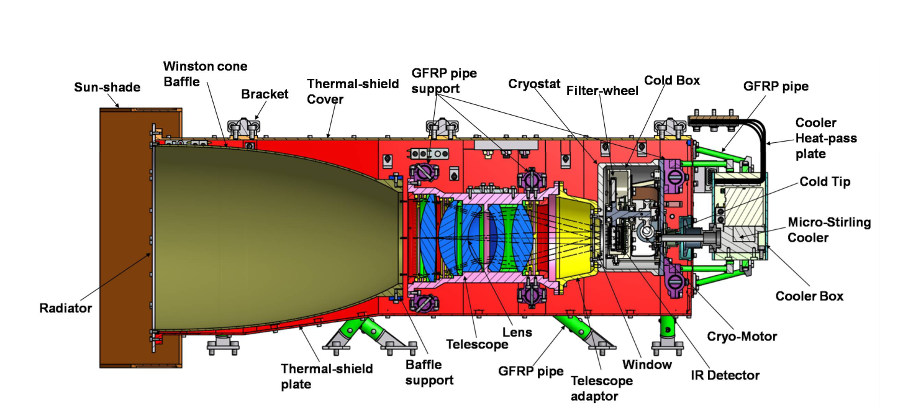}
\caption{
Mechanical configuration of MIRIS in a cross-sectional view. 
Adopted from \cite{10.1117...12.856364}. 
}\label{fig_MIRIS_2}
\end{figure*}

Astronomical observation becomes practically possible when the infrared sensor is cooled down to cryogenic temperature around 90~K.  The MIRIS telescope system was designed to cool passively to 200~K utilizing the low temperature of the deep space.  The sensor module in the cryostat was further actively cooled down to 90~K using a micro cooler (Ricor K508 model).  Thermal simulations were performed considering all relevant components of MIRIS to examine the passive cooling concept, and proper operation of the sensor temperature was finally approved using the thermal-vacuum chamber during the laboratory tests \cite{2012JASS...29..305L}. 
It was found that cooling the MIRIS telescope system down to 200~K would require approximately 7 days by passive cooling technique in space environment, depending on orbital altitudes.  The infrared sensor in the cryostat reached 90~K in approximately 10 hours by the active cooling of the micro stirling cooler.  The radiator was designed to reduce the heat load on MIRIS using a special material applied on an aluminum surface for higher emissivity, and a Winston cone baffle \cite{1970JOSA...60..245W} 
was employed to minimize the infrared radiation inflow from the telescope exterior \cite{10.1117...12.856364}. 
Thirty layers of Multi-Layer Insulation (MLI) wrapped the thermal shield box to decrease thermal loading on the telescope and to limit  radiation inflow from the exterior of MIRIS. The MIRIS opto-mechanical layout is shown in Fig. \ref{fig_MIRIS_2} 
and specification is summarized in Table \ref{MIRIS_table1}.  

Two electronics subsystems were designed to control the interface with spacecraft bus (E-Box) and cryogenic dewar with the infrared sensor (cold electronics).  The E-Box is independently located inside the MIRIS system, consisting of seven boards handling power conversions, data and commands interface, house-keeping monitoring, filter wheels control, ADC conversion, and other necessary electronic subcontrols.  These boards are controlled by a single chip FPGA (Field Programmable Gate Array) which contains most of the control logics, timings of driving waveforms for the clocks of the infrared sensor, ADC control, and necessary communication signals.  The cold electronics was developed for the interface with infrared sensor to preamplify the output signals and it was located inside the cryogenic dewar.  The PICNIC 256$\times$256 sensor has four independent output signals, and to operate a pixel every 8 $\mu$s, fast clock was selected at 125~kHz.  The critical performance for the electronics is the amount of dark noise generated from the variation of thermal electrons within the sensor depending on its temperature, in contrast to real signal from incoming photons.   The gain and readout noise of the system were measured around 10~e$^{-}$/ADU and 45~e$^{-}$, respectively.

MIRIS engineering model was assembled for shock test to examine its tolerance, particularly during launch, using a shock simulator at KARI.  Unfortunately, early opto-mechanical design of the engineering model failed in the shock test with serious damage including that of three lenses among five in the telescope.  After analyzing the shock response spectrum, it was concluded that the failure mode was a resonance at 100~Hz between MIRIS and flexure in response to an acceleration of about 100~G.  To address this failure mode, the project was delayed for 6 months, during which the flight model was redesigned with flexure preloads and spacer mounting designs were improved.  During the final shock test of the flight model with improved design, no damage or alteration was detected for shock levels of up to 200~G.  The optical performance of MIRIS was verified by using a profiler before and after the shock tests \cite{2014PASP..126..853H}. 

In parallel with system design, the data reduction software package was also developed, which included bad pixel masking, linearity correction, differentiation, flat-field correction, astrometry correction, and image stacking for data processing.  Integrated with MIRIS, STSAT-3 was successfully launched by the Dnepr Russian rocket on November 2013 and arrived at the Sun-synchronous orbits with an altitude of approximately 600~km.  The first light was obtained for M42 (Orion Nebula) and Rosette Nebula, which indicated correct cryogenic operation of the infrared sensor at around 90K and the telescope temperature around 200K by the passive cooling.  From March 2014, MIRIS observed the Galactic plane with Pa$\alpha$ line and fluctuations in the cosmic near-infrared background radiation as the main science mission on a regular basis until the lifetime of the stirling cooler.

\subsection{Science Contributions}

MIRIS Pa$\alpha$ Galactic Plane Survey (MIPAPS) covered the entire Galactic plane within the latitude range of $-3\degr\la\it b\la$$+3\degr$ at Pa$\alpha$ (1.87 $\mu$m).   The first result of the MIPAPS data extracted from the longitude range of $\ell$ = 96.5$^\circ$$ \sim$116.3$^\circ$ was obtained and the quality and scientific potential of the data were demonstrated by comparing them with H$\alpha$ maps obtained from the Isaac Newton Telescope (INT) Photometric H$\alpha$ Survey (IPHAS) data.  The Pa$\alpha$ and H$\alpha$ images were visually examined for the 31 ``Known'', 71 ``Candidate", 18 ``Group", and 92 ``Radio Quiet'' sources in the  $\it{WISE}$ (Wide-field Infrared Survey Explorer) H~II region catalog.  The examination newly identified 90 H~II region candidates (54 ``Candidate'', 18 ``Group'', and 18 ``Radio Quiet'' sources) as definite H II regions with the detection of the Pa$\alpha$ and/or H$\alpha$ recombination lines, out of which 53 H~II regions (39 ``Candidate'', 12 ``Group'', and 2 ``Radio Quiet'' sources) were detected at Pa$\alpha$.  The Pa$\alpha$ emission line has an advantage of being less attenuated compared to the H$\alpha$ line, in particular in regions with higher extinction such as the inner Galaxy.  Therefore, we expect that the entire data set of MIPAPS will allow us to identify more H~II region candidates as true H~II regions. It is estimated that the entire MIPAPS Pa$\alpha$ data will allow us to identify more than 1,600 $\it{WISE}$ H~II region candidates as true H~II regions \cite{2018ApJS..238...28K}. 

The examination also detected new 29 extended and 18 point-like sources at Pa$\alpha$ which were not included in the $\it{WISE}$ H~II region catalog. Their counterparts at H$\alpha$ were also found for all of the 29 extended sources and 16 point-like sources. In SIMBAD, it was found that 16 Pa$\alpha$ extended sources are associated with known H~II regions, and 18 Pa$\alpha$ point-like sources are due to 3 planetary nebulae and 15 emission-line stars which include 6 Wolf-Rayet and 2 Herbig Ae/Be stars.  Out of the other 13 Pa$\alpha$ extended sources with no known counterparts, 9 Pa$\alpha$ extended sources are considered to be newly identified H~II regions based on their isolated circular morphologies at Pa$\alpha$ and H$\alpha$.   The $\it{E(B-V)}$ color excesses for 62 $\it{WISE}~$H~II regions (22 ``Known'', 30 ``Candidate'', 9 ``Group'', and 1 ``Radio Quiet'' sources) and 16 MIPAPS Pa$\alpha$ extended sources were estimated by measuring the MIPAPS Pa$\alpha$ and IPHAS H$\alpha$ total fluxes.  It was found that the $\it{E(B-V)}$ values are basically in proportion to the distances to the sources, which is consistent with the fact that the more distant sources are commonly more attenuated due to the larger amount of interstellar dust. Two exceptional sources, which have relatively high $\it{E(B-V)}$ (in comparison with distance), could be young ultracompact H~II regions still embedded in dense molecular clouds. The $\it{E(B-V)}$ values were found to be similar to those estimated from point stars associated with H~II regions if we assume that the ratio of N~II to H$\alpha$ is 0.2 for the H~II regions.

\begin{figure*}[t]
\includegraphics[width=17cm]{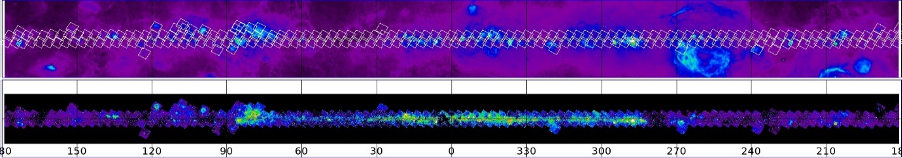}
\caption{
The upper image is the full-sky H$\alpha$  Map by Finkbeiner \cite{2003ApJS..146..407F}, and the below image is the MIRIS  Pa$\alpha$ survey image (continuum-subtracted) along the entire Galactic Plane (Milky Way) in Kim et al.~\cite{2018ApJS..238...28K} where the rectangles are the 235 fields observed by MIRIS and the numbers below are the galactic longitudes.
}\label{fig_MIRIS_3}
\end{figure*}

The observed Pa$\alpha$ total fluxes and the estimated $\it{E(B-V)}$ values were also used to calculate the total Lyman continuum luminosities of H~II regions.  The total Lyman continuum luminosity calculated for each H~II region can be used to constrain either the distance to the source or the spectral type(s) of its ionizing star(s) if we know one of these two. The total luminosities for 9 H~II regions are found to simultaneously agree with their known distances and the spectral types of their known ionizing stars.  Utilizing the MIPAPS Pa$\alpha$ and IPHAS H$\alpha$ images, we obtained an $\it{E(B-V)}$ map for the entire region of the H~II region Sh2-131 which is one of the largest H~II regions in the Cepheus region with an angular size of $\sim 2.5^{\circ}$.  
The $\it{E(B-V)}$ map shows a lot of high-extinction filamentary features in the northwest region which suggest the existence of foreground dust clouds in the region.  Fig. \ref{fig_MIRIS_3} shows the whole observed MIPAPS data, and further details were extensively described in Kim et al.~\cite{2018ApJS..238...28K}.

The analyzed MIPAPS data in the study mentioned above is less than 10\% of the entire Galactic longitudes, and the rest of them requires further analysis.   Apart from the MIPAPS data using Pa$\alpha$ images, the observed data with the broadband filters $\it{I, H}$ for high Galactic latitude clouds were analyzed to investigate DGL.  More details were described in Onishi et al.~\cite{2018PASJ...70...76O}.


\section{GALEX Mission}\label{GALEX_sec}


GALEX was a NASA-led UV imaging and spectroscopic survey mission \cite{2005ApJ...619L...1M}. 
With a 50 cm diameter primary mirror in a Richey-Chretien F/6 configuration, the GALEX telescope was designed for a wide field of view of 1.2 degrees and optimized for the UV observations in wavelengths from 135 to 280 nanometers. Yonsei university was one of the two international partners in this mission. As such, the GALEX project was the first official case of Korea-NASA cooperation on major space science program. From 1998 to the launch of GALEX in 2003, Yonsei scientists have participated in the GALEX UV instrument calibration and the pipeline software development. They have also independently developed several key softwares for the GALEX science operation and data analysis (SODA). Since its successful launch in 2003, the GALEX mission has mainly conducted its key science project that measures UV fluxes from galaxies to investigate star formation history in galaxies across $80\%$ of the age of the universe. In addition to this international collaborative project, Yonsei astronomers have also accomplished several milestone studies on the UV look-back time evolution of giant elliptical galaxies, the star formation history and supermassive black hole feedback in early-type galaxies, and the origin of super-helium-rich stars found in globular clusters and elliptical galaxies. The GALEX mission was decommissioned in 2013. Throughout the mission period, the GALEX science team has published more than one hundred papers mostly in the dedicated issues of the Astrophysical Journal Letters and the Astrophysical Journal Supplement Series, for which some significant fraction was led by Yonsei astronomers including two papers published in Nature and Science. 
   
   One of the two representative scientific discoveries made by Yonsei scientists in the GALEX mission was a successful detection of the look-back time evolution of the UV upturn phenomenon observed in normal early-type galaxies \cite{2005ApJ...619L.103L,2007ApJS..173..607R}. 
   For this project, the GALEX UV photometries of the brightest elliptical galaxies in twelve remote Abell clusters at moderate redshifts ($0.1 < z < 0.2$) were compared with the nearby giant elliptical galaxies in Fornax and Virgo clusters. Compared to their counterparts in nearby clusters, the far-UV flux of cluster giant elliptical galaxies at moderate redshifts fades rapidly with 2 Gyr of look-back time, and the observed pace in far-UV color evolution agrees reasonably well with the prediction from the Yonsei population synthesis models. This finding led to the conclusion that the long-standing puzzle of far-UV radiation from normal giant elliptical galaxies is hot helium-burning horizontal-branch stars and their progeny.  The other key achievement was the discovery of early-type galaxies with recent star formation and the suppression of star formation in massive early-type galaxies by feedback from supermassive black holes \cite{2005ApJ...619L.111Y,2006Natur.442..888S}. 
   Before this discovery from the GALEX mission, it was widely believed that early-type galaxies in general contain no young stars or signs of recent star formation. The GALEX UV survey of early-type galaxies clearly showed, however, that some significant fraction of these galaxies underwent recent star formation. This fraction is higher in less massive early-type galaxies with no signs of recent star formation among the most massive galaxies. It was shown that this result is related with the presence of supermassive black holes in the innermost central parts of galaxies. Supermassive black holes interact with their host galaxies by means of feedback which profoundly affects the evolution of host and its black hole. The unique GALEX observations of the UV emissions from massive early-type galaxies were used as constraint for the co-evolution of galaxies and black holes. An empirical relation was successfully derived for a critical black hole mass as a function of host mass above which the outflows from these black holes suppresss star formation in galaxies by heating and expelling the cold gas. This provided good evidence that the star formation in early-type galaxies is indeed regulated by feedback from supermassive black holes. 

   Korean participation in the GALEX project was successful mostly because it was science oriented and Yonsei astronomers played their own role in one of the key science goals of the mission. Stable long-term support from the Ministry of Science \& Technology of Korea was also an essential element for this success. Because of its relatively small aperture size, the GALEX instrument was best optimized for the survey purpose, and the detailed target based science programs were rather limited to nearby objects. In this respect, for Korean astrophysics community, the most ideal follow-up UV project would be to join a new generation UV-Visible space telescope project (e.g., post-Hubble Space Telescope mission) in the form of an international collaboration.

\section{AKARI}\label{AKARI_sec}


Infrared covers a wide range of wavelengths: from about 1 to 200 $\mu$m. 
AKARI was a Japanese infrared space  telescope  project with the original code name of ASTRO-F. After the successful launch in 2006, it was renamed to AKARI which means warm light in Japanese.

In December 1999, three Japanese astronomers (Toshio Matsumoto of Institute for Space and Astronautical Science [ISAS]\footnote{ISAS was an independent organization that belonged to the Ministry of Science, Technology and Education of Japan until 2005 when ISAS, National Aerospace Laboratory (NAL) and National Space Development Agency (NASDA) merged into Japan Aerospace Exploration Agency (JAXA). Currently, ISAS is a part of JAXA.}, Hiroshi Shibai of Nagoya University, and Munetaka Ueno of University of Tokyo) visited Department of Astronomy, Seoul National University (SNU) in order to explore a possibility of collaboration on the space mission AKARI by ISAS. Ueno spent some time at SNU  as a post-doctoral fellow and helped SNU for the construction of the first near-infrared CCD camera based on PtSi detector \cite{PtSi}. Matsumoto was the PI of the AKARI and  Shibai was in charge of the far infrared instrument of AKARI. 
The visit was arranged by Ueno as he was quite familiar with faculty members of the SNU astronomy. AKARI  was designed as a very ambitious mission that will cover almost all infrared wavelengths with imaging and spectroscopic capabilities. One of the main purposes is to carry out the all-sky survey in far infrared  from 50 to 200 $\mu$m with resolution and  sensitivity far better than those  of the Infrared Astronomical Satellite (IRAS) which was the first unbiased all-sky survey mission in infrared \cite{IRAS_Project}. Lunched in 1983 and operated for about 10 months,  IRAS revolutionized our understanding of the universe by detecting large number of star-burst galaxies and disks around main-sequence stars. It also  obtained the first images of the Milky Way's core. 

It was  originally planned to collaborate with Infrared Processing and Analysis Center (IPAC) located in the Caltech Campus in Pasadena for data analysis for the data analysis of AKARI, but IPAC did not have enough human resources to join the data analysis efforts of AKARI. IPAC was busy with the preparation of another important infrared mission, Space Infrared Telescope Facility (SIRTF) in US. SIRTF was eventually launched in August 2003 and renamed later as Spitzer Space Telescope \cite{Spitzer_Project}, after the great astronomer Lyman Spitzer Jr. who had promoted the concept of space telescopes in the 1940s. He was also one of the pioneers of the study of the interstellar medium. Eventually AKARI leadership wanted some help from the SNU astronomers for data analysis. Later, European institutes such as Imperial College London, University of Kent, Sussex University, Space Research Organization of the Netherlands (SRON) with the Kapteyn Astronomical Institute and European Space Agency joined the data analysis team.

After the visit of Japanese scientists in late 1999, the faculty members of SNU astronomy all agreed to take part in the AKARI project, but there are many practical things to solve. Although Japanese did not ask for any cash contribution, they wanted at least one senior astronomer to reside at ISAS together with young people such as graduate students and post-doctoral fellows. The senior member will be supported by ISAS through visiting professor program, but the cost and salary for the young people have to be paid by Korea.  Fortunately, graduate students could receive financial support through Brain Korea 21 (BK21) program which started in 1999. BK21 emphasized the importance of the international collaboration, and sending students abroad through this program was highly encouraged. At SNU, there was another problem. The  students who have not finished the course work had to earn course credits while they were away. This problem was solved through the agreement between SNU and University of Tokyo on credit exchange.

Hyung Mok Lee was decided as the first senior member to be stationed at ISAS from September 2000, and SNU granted leave of absence status for him for one year. He went to ISAS with three graduate students: Woong-Seob Jeong, Jungjoo Sohn, and Sam Kim. The first task of the Korean team was construction of the simulation software that mimics the instrumental characteristics and the observing scenarios as close as possible for far infrared scanning observations. The scope of the works of the Korean team has expanded to the development of the scientific programs, analysis of the laboratory data for the detectors, and calibration scheme. The senior members who stayed at ISAS for AKARI after Hyung Mok Lee include Soojong Pak, Seung Soo Hong, Bon-Chul Koo, Young-Sun Park and Suk Minn Kwon (Gangwon National University). Graduate students have been continuously supported by the BK21 program. 

\subsection{Overview of the AKARI Mission}

\begin{figure}
\includegraphics[width=9.0cm]{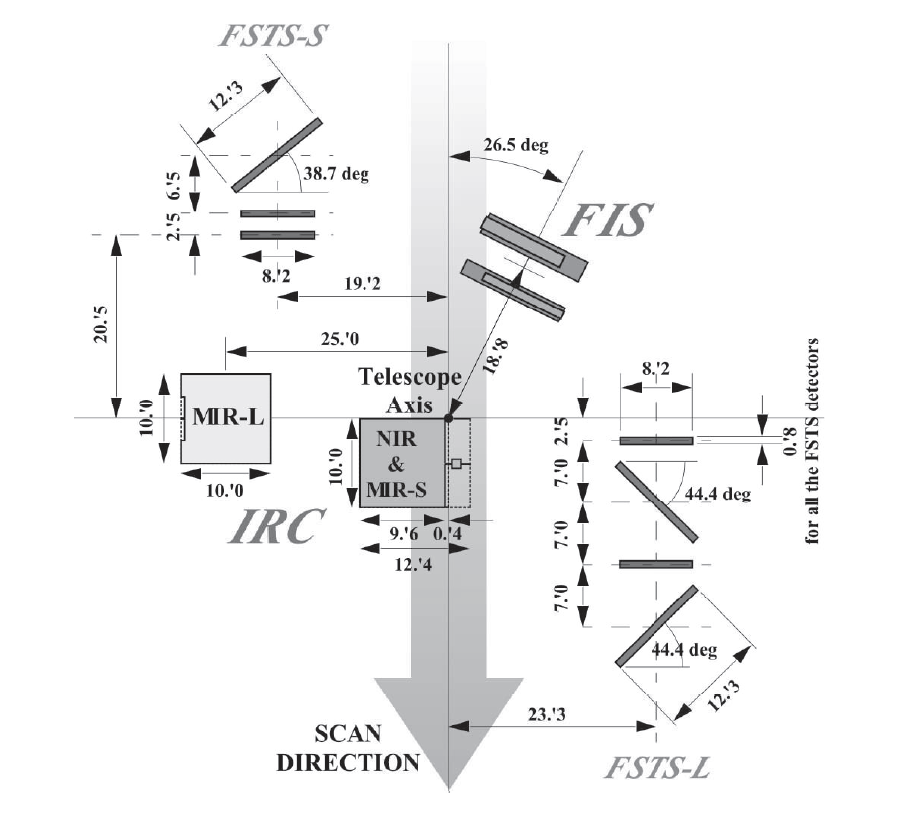}
\caption{Focal plane layout of AKARI projected on the sky. }
\label{fp_inst}
\end{figure}

\begin{figure}
\includegraphics[width=8.0cm]{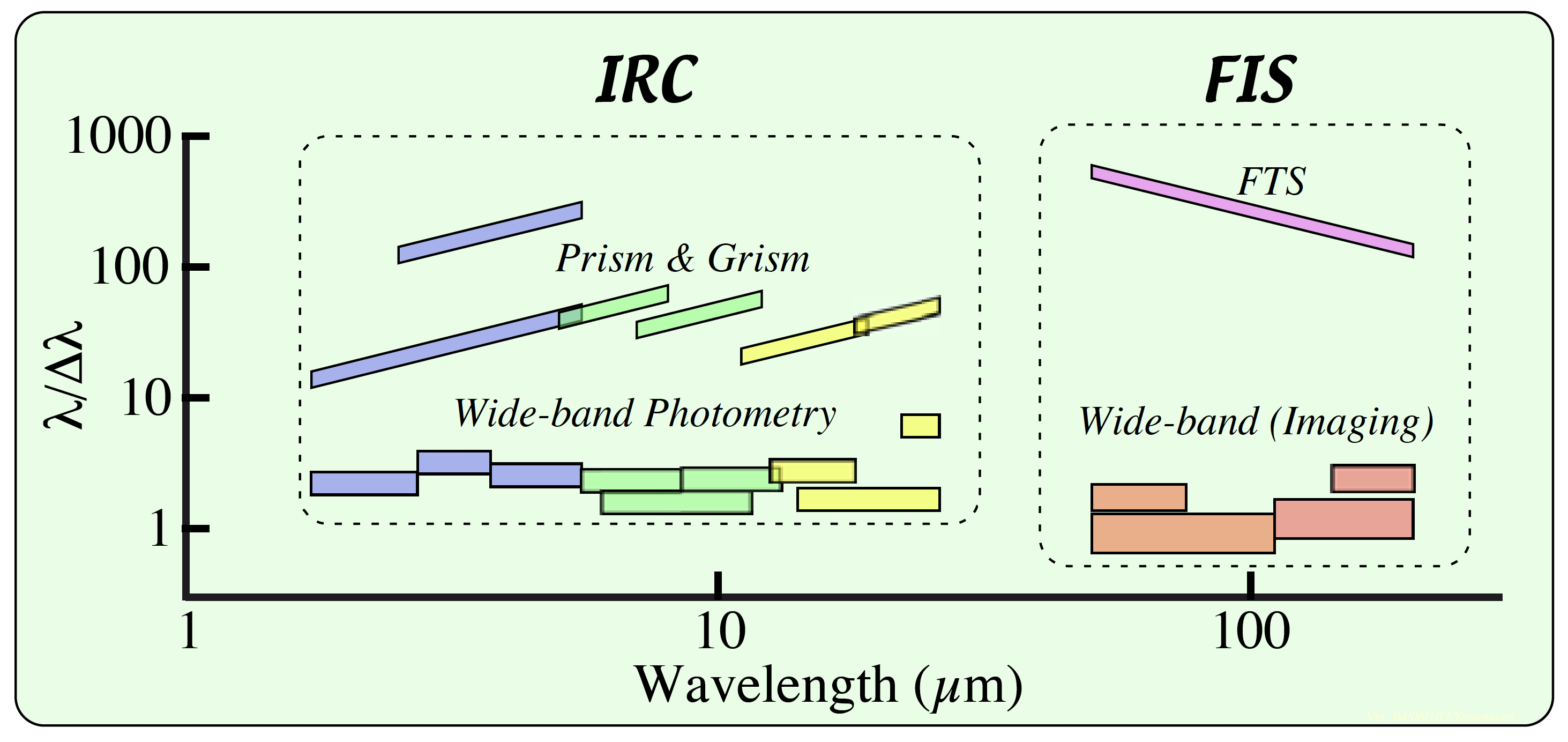}
\caption{Wavelength coverage and spectral resolution of the focal plane instruments of AKARI}
\label{akari-fig1}
\end{figure}

AKARI was designed to carry out the all-sky survey at far infrared and pointed observations in near and mid-infrared \cite{overview}. The telescope is a   Ritchey-Chretien  system with 68.5 cm main mirror made of silicon carbide (SiC). There are two focal-plane instruments. (1) Far Infrared Surveyor (FIS, \cite{FIS}) with two-dimensional detector arrays of Germanium doped Gallium (Ge:Ga) (20$\times$2 for N60 [60 $\mu$m] and 20$\times$3 for WIDE-S  [90 $\mu$m]) and stressed Ge:Ga sensors  (15$\times$3 for Wide-L [140 $\mu$m] and 15$\times$2 for N160 [160 $\mu$m]). These sensors are also shared with the Fourier Transform Spectrometer (FTS). (2) Infrared Camera (IRC) \cite{IRC} consists of three channels, NIR, MIR-S, and MIR-L, which cover $1.8 - 5.5$ $\mu$m, $4.6 - 13.4$ $\mu$m, and $12.6 - 26.5$ $\mu$m, respectively. The sensors for IRC are InSb 512$\times$412 array for NIR, and 256$\times$256 Si:As arrays for MIR-S and MIR-L. Each channel has three broad-band filters so that IRC has total nine bands for imaging. IRC also had  grism/prism spectrometers. The layout of the focal plane instruments projected on the sky, with focal plane star sensors (FSTS-S and FSTS-L), is  shown in Fig. {\ref{fp_inst}}. The wavelength coverage and resolution of the FIS and IRC are shown in Fig. {\ref{akari-fig1}}.  The entire telescope is cooled to about 6~K in order to suppress the thermal emission. The wavelength coverage of AKARI was much wider than IRAS which had  4 wide bands at 12, 25, 60 and 100 $\mu$m.  The spectroscopic capability of the IRAS was limited to 8  - 23 $\mu$m only at low resolution at $\lambda/\Delta\lambda = 14 - 35$.

\begin{figure}
\includegraphics[width=8.0cm]{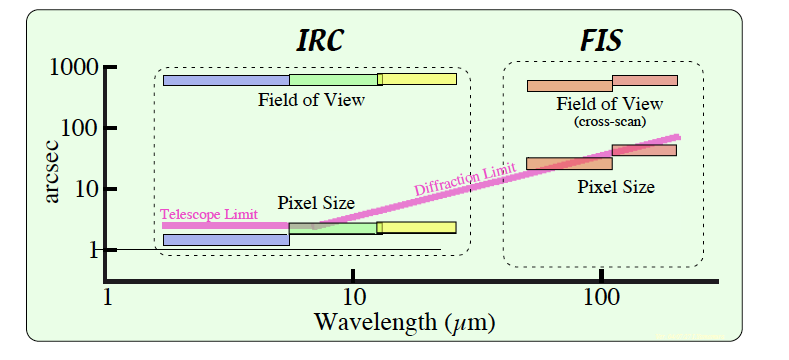}
\caption{Pixel size, field of view, and the angular resolution of the AKARI's instruments (Figure credit: ASTRO-F Observer's Manual Version 2.0)} 
\label{akari-fig2}
\end{figure}

The angular resolution of AKARI is much better than that of IRAS, thanks to the technological advances in infrared detectors. The size of the telescope aperture of IRAS was 57 cm which is only slightly smaller than that of AKARI, but the detector pixels of IRAS were very large and the angular resolution was mostly determined by the pixel sizes. For the case of AKARI, the pixel sizes of the detectors are comparable to the  diffraction limit of the telescope, except for near-infrared band. In near-infrared at $\lambda  \lesssim 7 ~\mu$m, the surface accuracy of the main mirror was not good enough to reach the diffraction limit. See Fig. \ref{akari-fig2} for actual angular resolution as a function of wavelength.

\begin{figure}
\includegraphics[width=8.0cm]{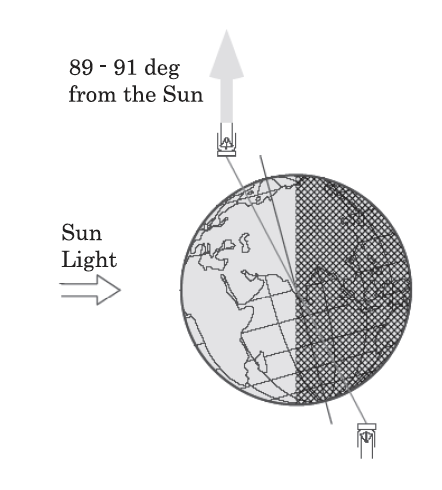}
\caption{Altitude of AKARI relative to the Sun and the Earth} 
\label{fig_orbit}
\end{figure}

AKARI was put on Sun-synchronous polar orbit at an altitude of about 700 km above the surface of the Earth with inclination angle of the  orbital plane relative to the sunlight being 98$^\circ$. The sunlight is shielded by the sun shield panel and the range of the allowed direction of the telescope was between 89$^\circ$ and 91$^\circ$ relative to the direction to the Sun and opposite to the Earth (see Fig. \ref{fig_orbit}). There were several modes of the observations: continuous scan at the speed of 3.6 arcmin per second for FIS all-sky survey, slow scan at the speed of  $\sim$ 30 arcseconds per second for higher sensitivity observations, and pointed observations. For the all-sky survey observations, the circular strip of about 8.2 arcmin  width is scanned each orbit.  Both North Ecliptic Pole (NEP) and South Ecliptic Pole (SEP) are visited while there is about 4.2 arcmin shift along the Ecliptic Equator in each orbit. The orbital period was about 100 minutes and  the entire sky can be observed in 6 months at least twice.  For IRC, the pointed observations with the maximum integration time of 10 minutes  are employed. 

The expected mission life under cryogenic condition was about 550 days so that the all-sky survey could be executed more than twice. Unfortunately an  unexpected problem of the Sun aspect sensor for crude altitude control forced the delay of the opening of the aperture lid. The liquid helium has to be consumed in order to keep the telescope at low temperature without radiative cooling for more than one month. The aperture lid was eventually ejected on April 13, 2006 after which performance verification phase started. The actual observations started on May 6, 2006 and lasted until August 26, 2007 on which the liquid helium ran out.  Additional observations in near-infrared only were carried out with mechanical coolers from June 2008, after the readjustment of the focus, assessment of the near infrared performance, and changing of the software for the downlink of the data were made. AKARI stopped Phase 3 on February 15, 2010 when the detector temperature became too high for observations due to the degradation of the cryocoolers. 


\subsection{Observational Programs}

The observations were carried out in three different phases. The first 6 months of the observation (Phase 1)  starting from May 8, 2006 were mostly dedicated to the all-sky survey with some pointed observations at high ecliptic latitudes. The Phase 2 observation, which started on November 10, 2006 and lasted until August 26, 2007 when the liquid helium ran out, was shared between pointed observations and supplemental all-sky survey. After the helium boiled off, near-infrared observations with cryocoolers only were done from June 8, 2008 until February 2010 (Phase 3). 

There were three types of the observing programs: Large Area Surveys (LAS), Mission Programs (MP), and Open-Time (OT) Programs. Both LAS and MP aimed to obtain maximum science output by dedicating large fraction of the telescope time. The most important LAS was the all-sky survey\cite{All_Sky1,All_Sky2}.  There were two additional LASs:  North Ecliptic Pole (NEP) Survey \cite{NEP_Survey} and Large Magellanic Cloud (LMC) Survey \cite{LMC_Survey}  that utilized the nature of the polar orbit. LMC is very close to the SEP.  While the all-sky survey is carried out by the scanning observations, NEP and LMC surveys are performed mostly with pointed observations.  

Fifteen MPs covering a wide range of objects, from solar system objects to distant galaxies, were approved before the launch of AKARI with detailed list of observational targets. After fixing LASs and MPs, the remaining telescope time was given to OT programs which were open to  Japanese, Korean and European astronomical communities. About 30\% of the Phase 2 observations were allocated for OT programs.  There were separate calls for MPs and OTs for the Phase 3 observations.

AKARI made about 5500 pointings of  imaging and spectroscopic observations with FIS and IRC during Phases 1 and 2, and 12,600 pointing observations during the Phase 3 (near-infrared only), in addition to the all-sky survey observations \cite{Nakagawa_PKAS}. Initially, the all-sky survey was planned to be done only in far-infrared channels, but surveys in two mid-infrared channels at 9 and 18 $\mu$m were  also done since the data could  be  simultaneously collected in IRC channels during the scanning observations \cite{MIR_ASS}. Only two rows of the MIR-S and MIR-L detectors were used for the all-sky survey observations because of the limitations in data downlink. 

\subsection{Data Products of Large Area Surveys}

The all-sky survey covered more than 96\% of the entire sky at 6 wavebands: 9, 18, 65, 90, 140, and 160 $\mu$m \cite{Nakagawa_PKAS}.  The data products of the all-sky survey include AKARI All-Sky Survey Bright Source Catalogue \cite{FIS_BSC1,FIS_BSC2} and AKARI IRC All-Sky Survey Catalogue\footnote{\url{https://irsa.ipac.caltech.edu/data/AKARI/documentation/AKARI-IRCPSCV1RN.pdf}}. 
All sky images at far infrared \cite{All_Sky1,All_Sky2} and at mid-infrared \cite{MIR_ASS} are also available. The FIS Bright Source Catalogue contains about 427,071 sources and the IRC All-Sky Bright Source Catalogue contains 870,974 sources. For comparison, IRAS point source catalogue contains about 250,000 sources. From the LMC Survey, both LMC Point Source Catalogue \cite{LMC_PSC}  and  Near-IR Spectral Catalogue are available. The NEP Survey Program provides NEP-Deep Field Point Source Catalogue \cite{NEP_Deep_MIR_PSC} and NEP-Wide Field Point Source Catalogue \cite{NEP_Wide_PSC}.  These data can be downloaded from  Data Archives and Transmission System (DARTS) of the ISAS/JAXA\footnote{\url{http://darts.isas.jaxa.jp/astro/akari/cas/index.html}}. In addition, data contributed by the users based on  other programs are also available from the same site.

\subsection{Contribution by the Korean Group}

Although the collaboration was initiated by the faculty members at SNU, a large number of Korean astronomers were able to be involved in the development and execution of all observing programs. For the FIS All-Sky Survey, the main contribution of the Korean team was the development of the simulation software that allowed the performance test and scheduling of the observation \cite{fisvi}. Contributions from the Korean group in the efforts of early data analysis include 
quick analysis on the early survey of NEP--Deep field \cite{Early_NEP}, comparison of FIS scanning observation data with IRAS \cite{Early_FIS}, and detection of the supernova remnant from the  Small Magellanic Cloud (SMC) \cite{SNR_SMC}. 

\begin{figure}
\includegraphics[width=8.0cm]{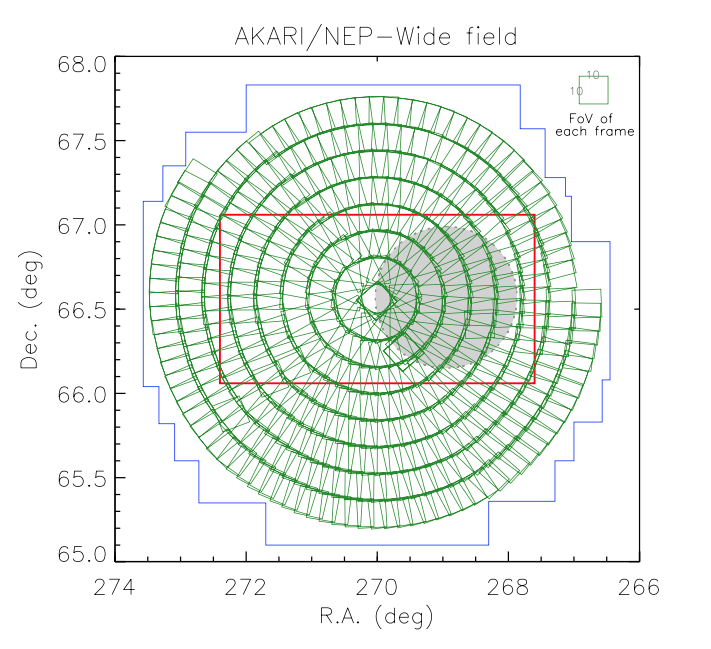}
\caption{Survey areas of NEP-Wide (large circle)  and Deep (small shadowed circle on the right). Survey areas of preflight optical survey with CFHT (red box) \cite{CFHT_NEP} and optical survey with Maidanak Observatory (blue line) \cite{Maidanak_Survey} are also shown. Each green box represents the IRC pointing field.}
\label{NEP_Field}
\end{figure}

Among three LASs, the Korean group was most deeply involved in the NEP and LMC Surveys. NEP Survey was composed of two programs: NEP--Wide  and NEP--Deep \cite{NEP_Survey}. NEP-Wide field is a circular shaped area of  5.4 square degrees centered on the NEP, and NEP--Deep is a much smaller circular area of about 0.5 square degrees inside the NEP--Wide field.  NEP--Wide and NEP--Deep fields were observed with 446 and 266 pointings. These surveys were designed to be complementary to each other: the Wide Survey was intended to cover a large area in order to avoid cosmic variance while Deep Survey was designed to reach faint and distant objects. The  areas of the Wide and Deep Surveys are shown in Fig. \ref{NEP_Field}. The Korean Group has obtained deep optical images for the large fraction of the NEP-Wide field using CFH12K camera (red box in Fig. \ref{NEP_Field}) attached to Canada-France-Hawaii Telescope (CFHT) and published optical source catalogue \cite{CFHT_NEP}.  The Korean Group led the analysis of the NEP-Wide Survey data and followup studies, including the production of the point source catalogue of NEP-Wide field \cite{NEP_Wide_PSC},  optical survey of the entire NEP-Wide field using 1.5 m telescope at Maidanak Observatory in Uzbekistan \cite{Maidanak_Survey},  redshift survey of selected galaxies \cite{NEP_redshift},  and submillimeter wavelength survey using James Clerk Maxwell Telescope (JCMT) \cite{JCMT_NEP}. Korean Group has produced the catalogue of supernova remnants by analyzing the LMC Survey data\cite{SNR_LMC}.

During the Phase 3 observation period, Korean Group obtained a large number of near-infrared spectra for quasars and active galactic nuclei (AGNs)  and studied the evolution of supermassive black holes \cite{Quasar_BH} based on the mass estimates with the same spectral line (i.e., H$_\alpha$ for a wide range of redshifts). A few other examples of the Korean contribution include the analysis of the Zodiacal light in infrared \cite{IRC_Zodiacal}, study of supernova remnants in LMC \cite{SNR_LMC}, infrared properties of galaxies in clusters of galaxies \cite{CLEVL}, and near-infrared background light \cite{EBL}.  

The contribution listed above is only a small fraction of what Korean group has done with the data obtained with AKARI. The AKARI data are still a very useful resource for wide areas of astrophysical research.  In the absence of any near-term far-infrared space mission, the AKARI data will remain a unique resource.  

\subsection{Heritages of the AKARI Project to Korean Community}

Korean group collaborated in the AKARI project without any hardware contribution mainly because there was no expertise in space infrared instruments in Korea at that time. The collaborative works such as software development, analyzing and modeling the experimental data, and data analysis have been done by several students. Because the duration of the space project could be longer than the graduate school period, the contributing students sometimes could not use the scientific data for their PhD theses. Many of the early contributors therefore had to choose different topics for their theses. The number of PhD theses from SNU based on the AKARI observations was eight.

However, there remain substantial heritages from their experiences with the AKARI project in Korea. The Korean members of AKARI started community-wide workshops by forming an Infrared Astronomy Working Group (IRWG) in 2001. Initially, IRWG tried to promote infrared astronomy in order to let the Korean astronomers become more aware of the availability of observing time through OT proposal.  Also, IRWG was very keen to develop infrared projects in Korea after AKARI and  invited people in various fields including industry. 

In Japan, the infrared community was preparing for a very ambitious infrared project called SPICA which stands for the SPace Infrared telescope for Cosmology and Astrophysics. The first international workshop was held in 2000 at ISAS where a few Korean astronomers also participated. It was planned to be  placed in the L2 halo orbit which is about 1.5 million km from the Earth in order to be free from the heating by the Earth. The entire telescope will be cooled to around 6 K so that observations in far infrared become possible. Initially, it was assumed that SPICA would cover from near-infrared to far infrared, but later the main focus became mid-infrared and far infrared, since the James Webb Space Telescope (JWST) would extend its coverage to almost 20 $\mu$m. The telescope aperture was 3.5 m in the initial design.

Korea was also invited to be part of the SPICA project with contribution to the focal-plane instrument. Since Korea did not have much experience in mid and far infrared, Korean members proposed to provide Focal Plane Camera (FPC) for near-infrared camera in the focal plane that can be used for accurate attitude control and pointing reconstruction as well as scientific instrument. Although near-infrared was not the core wavelength of the  SPICA, the sensor for the attitude control was required and the Korean proposal was to make powerful enough for scientific purposes. Because of the large aperture of SPICA, the near-infrared data automatically taken by the FPC during the entire mission period  could be very useful resources with excellent sensitivity and  angular resolution.  Possible science programs with the FPC include the investigation of near-infrared cosmic background radiation.

However, the Japanese space agency JAXA decided SPICA to be a joint project with ESA  and excluded Asian partners such as Korea and Taiwan. Currently SPICA is one of the candidates for the next Medium Class Mission 5 of the Cosmic Vision program in ESA. If selected, it will be launched in 2032. Also the diameter of the telescope was reduced to 2.5~m during the restructuring process.

The experience in the AKARI collaboration led to near-infrared project such as KASI Near Infrared Camera System (KASINICS) which is a near-infrared imaging camera for ground based telescope, in particular, 1.8~m Doyak Telescope at Bohyunsan, MIRIS on-board  the STSAT-3, and Near-infrared Imaging Spectrometer for Star formation history (NISS) on-board  NEXTSAT-1. Furthermore, the concept of near-infrared spectroscopy using linear variable filter (LVF) for NISS was extended to SPHEREx which was selected as a MIDEX by NASA in 2019. KASI is the only foreign partner of the SPHEREx project. MIRIS, and NISS/SPHEREx are covered in Sections \ref{MIRIS_sec} and \ref{NISS_SPHEREx_sec}, respectively.

\section{UFFO-Pathfinder: Gamma-Ray Bursts Mission}\label{UFFO_sec}


\subsection{Observation of Prompt Photons from Gamma-ray Bursts}

Gamma-ray bursts (GRBs) are the most energetic explosions in the universe and are not only extremely interesting phenomena in astrophysics but also important for multi-messenger astrophysics with gravitational waves and cosmic-ray particles \cite{2017SSRv..207...63W}. 
GRBs happen every day, in any place in the sky and last only a few tens of seconds. However, its origin and mechanism are still unknown yet. GRBs are detected first with gamma-rays or X-rays and followed typically by UV/optical and radio afterglows, which implies that what causes GRBs emits photons in a great range of  wavelengths. UV/optical emission at early times is believed to originate from external shocks which are predicted to have a smooth, monotonic rise \cite{2004RvMP...76.1143P}. 
Observation of an early-time UV/optical light curve that more closely resembles a gamma-ray light curve  (e.g., jagged and with multiple peaks) would clearly indicate the presence of prompt optical emission produced by internal shock. 

NASA’s Swift  
space mission, launched in 2004 \cite{2004ApJ...611.1005G}, observed hundreds of GRBs and measured GRB UV/optical light curves. However, Swift has such a hardware limitation that it has to rotate the entire spacecraft in order to observe the UV/optical light of GRB by using the UV/optical telescope onboard after acquiring the first position of the GRB with the burst alert telescope (BAT: onboard as well). As a result, Swift has an intrinsic gap of about 60 seconds before the optical measurement \cite{2005SSRv..120...95R}. 
Although two or three observations of an early-time UV/optical light curve are available so far, they are not sufficient to study the burst mechanism and correlation between prompt emission and early UV/optical emission. Sub-minute measurements would be required to learn more about such prompt emission signatures in the early UV/optical emissions. 

\begin{figure}
\includegraphics[width=8.8cm]{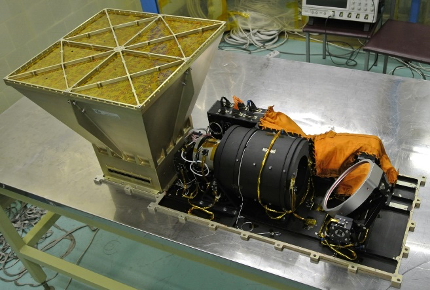}
\caption{
Fabricated UFFO-pathfinder flight model. The left gold color structure is UBAT and right is SMT without housing.
}\label{UFFO_fig1}
\end{figure}

\subsection{Construction of the UFFO-pathfinder Payload onboard the Lomonosov Spacecraft}

UFFO was proposed by Il Park in 2009 for precise observation and identification of early UV/optical photons from GRBs 
\cite{2009arXiv0912.0773P,2013NJPh...15b3031P}. 
It was an international joint project in which about 17 universities/institutions from 9 countries (Korea, Russia, Taiwan, Spain, Denmark, Norway, France, Poland, and USA) participated. A new method was proposed to measure GRB UV/optical light curves within seconds after the gamma-ray signal. The key idea was to rotate only the mirror of the telescope instead of rotating the entire satellite after receiving the GRB position. Based on this concept, we designed, developed, and fabricated two types of telescope, the Slewing Mirror Telescope (SMT) and the UFFO Burst Alert and trigger Telescope (UBAT). We describe them in detail below. 

\begin{figure}
\includegraphics[width=8.8cm]{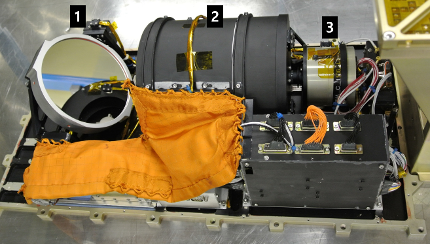}
\caption{
Integrated SMT. (1) Slewing mirror system consisting of 2-axis gimbal mirror; (2) 10~cm aperture Ritchey-Chrétien telescope; (3) ICCD
}\label{UFFO_fig2}
\end{figure}

SMT is the key of UFFO-pathfinder and aims to observe UV/optical photons by fast moving the mirror to the position of the GRB received from UBAT \cite{2018SSRv..214...16J}. 
%
SMT consists of a slewing mirror system, a Ritchey-Chretien telescope (RC telescope) with a 10~cm diameter aperture \cite{2013OExpr..21.2263J}, 
an Intensified Charge-Coupled Device (ICCD) \cite{2013JInst...8P7012K},  
and electronics. The slewing mirror system includes a 15~cm diameter flat mirror and a two-axis gimbal stage which is driven by two stepping motors. The mirror moves up to $\pm 35 \degr$ 
which is equivalent to UBAT’s half-coded FOV of $70\degr \times 70\degr$. As shown in Figs. \ref{UFFO_fig1} and \ref{UFFO_fig2}, the slewing mirror system is placed in front of the RC telescope. The GRB UV/optical light is reflected in the slewing mirror, passes through the RC telescope, and enters ICCD. ICCD has a size of $256\times256$ pixels and the FOV of a pixel is 4$\rm''$$\times$4$\rm''$, so SMT has a FOV of 17$\rm'$$\times$17$\rm'$. The SMT observation time is 80 seconds and a total of 1700 frames of data is acquired. The exposure time of the first 880 frames is 20 ms, the exposure time of the subsequent 820 frames is 40 ms, and there is a 20 ms gap between each frame \cite{2017OExpr..2529143G}. 
The exposure time can be adjusted via the interface between the ground and satellite.

\begin{figure}
\includegraphics[width=8.8cm]{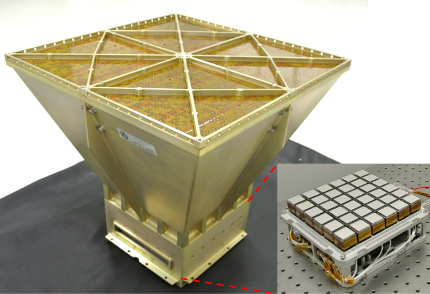}
\caption{
Integrated UBAT. The coded mask and hopper (left) and detector inside the structure (right bottom). 
}\label{UFFO_fig3}
\end{figure}

UBAT detects X-rays in the energy range of $10-200$ keV 
and triggers the detection of the GRB X-ray with a wide field of view. Its FOV is $90.4\degr \times 90.4\degr$ for partial coded and $70.4\degr \times 70.4\degr$ for half-coded, and the GRB localization accuracy is 10$\rm'$ for $>7\sigma$. UBAT consists of coded mask, hopper structure, detector which includes multi-anode photomultipliers (MAPMT) and YSO (Y2SiO5) crystal, and readout electronics (Fig. \ref{UFFO_fig3}). Coded mask is a technique to determine the direction of GRB by analyzing the pattern detected by the detector when the X-ray comes in. The UBAT coded mask has a pixel size of 5.67 mm $\times$ 5.67 mm and $44.5\%$ of pixels are randomly open while the rest are blocked by a 1~mm thick tungsten alloy. Hopper is made of 3~mm thick aluminum and 0.2~mm thick tungsten and surrounds four sides of the detector and coded mask to block the X-ray background from space. The detector has 2304 pixels in total area of $191.2~\mbox{cm}^2$ with $8\times8$ pixel MAPMTs and YSO crystals arranged in $6\times6$. We were the first to use a YSO crystal in a space X-ray detector. Readout electronics is composed of analog and digital electronics. There are 4 SPACIROC ASICs in an analog electronics and an ASIC chip is connected with a MAPMT, i.e., 64 channels. It performs photon counting and energy measurement of X-ray. Digital electronics has two field programmable gate arrays (FPGA): one serves to process and store the X-ray data from GRB and the other to perform operation for GRB trigger and to calculate the location of GRB.

The UFFO Data Acquisition system (UDAQ) is head of UFFO-pathfinder, controls the system operation, communicates between UBAT and SMT, interfaces with satellite, and acquires data. It is implemented in FPGA for fast processing speed and low power consumption, which is also the first to be used in space experiments.

After completing the fabrication of UFFO-Pathfinder, space environment test was performed at the Taiwan National Space Agency (NSPO) in August 2011. Before moving to Vostochny in 2016, the flight model was integrated with the Lomonosov \cite{2017SSRv..212.1705S}
satellite and successfully passed the space environment test at VNIIEM in Russia. 

\subsection{Flight Performance of UFFO-pathfinder}

The UFFO-pathfinder was mounted on the Lomonosov satellite, launched successfully from Vostochny, Russia on April 28, 2016, and operated in the Sun-synchronous orbit at the altitude of 550~km from the Earth. Unfortunately, UFFO-pathfinder was unable to observe GRBs due to a problem in the satellite’s power supply. Nonetheless, we acquired 13 sets of UV/visible and X-ray background data out of 19 calibration runs over a period of about 6 months in space, with a number of meaningful results.

We have confirmed that UBAT's 2304 channel detector operates successfully in space from the acquired X-ray background data. As proven in laboratory tests (M. B. Kim et al. in preparation), 
the X-ray energy is known to be up to 5 keV. It is the first time with UBAT to observe low energy X-ray by using YSO crystals in space.

\begin{figure}
\includegraphics[width=8.8cm]{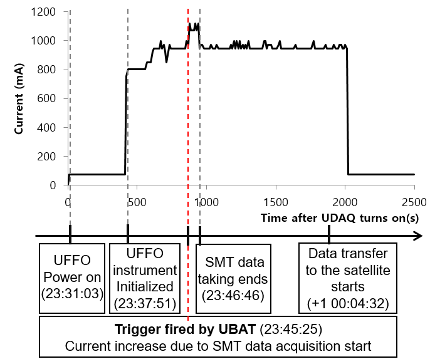}
\caption{
Run sequence with measurement and log of UFFO-pathfinder on Sep. 14, 2016. The time of trigger fired and SMT data taking starts are shown.
}\label{UFFO_fig4}
\end{figure}

Fig. \ref{UFFO_fig4} shows the change in power consumption over time in calibration running of the UFFO-pathfinder on Sep. 14, 2016. In this calibration, a fake trigger was created by lowering the trigger threshold to verify the behavior of instrument. We have confirmed that the UFFO-pathfinder operates in the same sequence we designed. After the UDAQ is powered on, the instrument is initialized. After that, when a trigger occurs, SMT starts data acquisition and it takes 82 seconds for the data taking to finish. Finally data transfer to the satellite. This result proved our main idea that SMT begins to collect UV/optical data less than 2~seconds after the trigger occurrence, i.e., the most rapid optical follow up observation in the world.

\begin{figure}
\includegraphics[width=8.8cm]{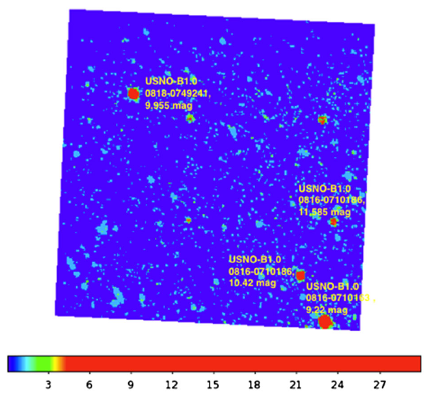}
\caption{The sky image of SMT observed on Sep. 8, 2016 during the SMT calibration on orbit. We used USNO-B catalog to check magnitude~\cite{2018SSRv..214...14P}.}\label{UFFO_fig5}
\end{figure}

Fig. \ref{UFFO_fig5} shows the image taken by SMT during the 7th calibration run of the UFFO-pathfinder on September 8, 2016. The exposure time was 0.6~s and the FOV is 17.7$\rm'$$\times$17.7$\rm´$. The center of the image is located at RA=22:46:46.56 and Dec=-0.8:14:31.20. Using the USNO-B catalog, we confirmed that the UV/optical sensitivity of SMT is about 19 magnitude for 100~s, as designed.

\subsection{UFFO-100: Next Generation GRB Mission}

\begin{figure}
\includegraphics[width=8.8cm]{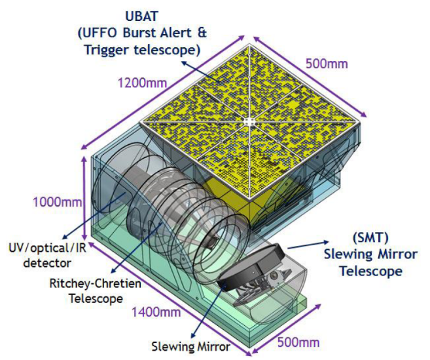}
\caption{
A rendering of UFFO-100 \cite{2013NJPh...15b3031P}
}\label{UFFO_fig6}
\end{figure}

Through the successful launch and space operation of the UFFO-pathfinder, the first mission of the UFFO project, we confirmed the realization of our ideas and the possibility of future GRB missions. The next mission of UFFO project, UFFO-100, is a large GRB observation payload with a total mass of 100~kg \cite{2013NJPh...15b3031P}. 
It will be able to observe the initial moment of GRB with higher sensitivity. UFFO-100 will consist of a $1024~\mbox{cm}^2$ X-ray trigger telescope using coded mask technique, a 30~cm aperture tracking telescope, polarimeter, and a near-infrared telescope (Fig. \ref{UFFO_fig6}). Mass of 100~kg and the 100~W power consumption are allocated for the location of the UFFO-100 device, which makes it possible to carry out all the planned scientific tasks.

The UFFO-100 will be placed onboard the Russian spacecraft GAMMA-400, which includes the main payload: the GAMMA-400 gamma telescope (Gamma-Astronomical Multifunctional Modular Equipment), the ART-X-ray telescope, and plasma detectors. The mass of the spacecraft is about 2500~kg and the spacecraft will be built on the already well-proven platform Navigator developed by Lavochkin. The target orbit of the spacecraft is initially an ellipse with a perigee of 3,000~km and an apogee of 300,000~km. After 6 months, the orbit becomes circular with a radius of 200,000~km. The inclination is $51.8\degr$. The expected launch date of the spacecraft is 2025.

\section{Infrared Spectro-Photometric Missions: NISS and SPHERE$\textsc{x}$}\label{NISS_SPHEREx_sec}



The infrared observations on the ground have been made at limited wavelengths due to high thermal background and strong atmospheric emissions. Since high throughput for diffuse objects (size $> 1\rm'$) even with a small telescope can be achievable in space by overcoming limitations of ground observations, the infrared space observations have been tried through small missions in Korea. After the termination of the STSAT series for the demonstration of space technologies and the research of space science, the new 100~kg-grade NEXTSAT has begun from 2012 to realize the miniature, modular and standard platform for small satellites even with low power consumption. Based upon the previous development of the infrared instrument onboard STSAT-3, the wide-field infrared imager MIRIS, KASI had successfully developed NISS as one of two scientific payloads onboard the first NEXTSAT \cite{Jeong18}. Based upon these heritages, KASI have participated in an international collaboration mission, the first all-sky infrared spectro-photometric surveyor SPHEREx \cite{Korngut18}, which was selected as a NASA MIDEX mission in February 2019. For the first time in Korea, KASI will contribute to the hardware development in a NASA Explorer program by providing the system calibration facility for SPHEREx. The spectro-photometry is a space observational technique that can obtain images and spectra simultaneously through the operation of LVF. Since a LVF should be installed very close to the sensor in a focal plane, there is no moving part and the instrument is operated in single observation mode, which can reduce any possible malfunctions in space. Table \ref{tab:spec_NISS-SPx} shows the specifications of NISS and SPHEREx. The SPHEREx mission dedicated to astrophysical research not only offers much better specifications but performs the all-sky survey. 

\begin{table}[h]
	\caption{Specifications of NISS and SPHEREx}
	\begin{ruledtabular}
		\begin{tabular}{lcc}
			Parameter & NISS & SPHEREx \\
			\colrule
			Field of view (deg) & 2$\times$2 & 3.5$\times$11.3  \\
			Aperture (cm) & 15 & 20  \\
			Spectral coverage ($\mu$m) & 0.95$-$2.5 & 0.75$-$5.0  \\
			Spatial resolution (arcsec) & 15 & 6  \\ 			
			Spectral resolution ($\lambda$/$\delta\lambda$) & 20 & 41$-$135  \\
			Sky coverage (deg$^{2}$) & $\sim$100 & all-sky  \\ 			 			
			Sensitivity (AB mag) & 17 & 19  \\ 			 			
		\end{tabular}
	\end{ruledtabular}
	\label{tab:spec_NISS-SPx}
\end{table}

\subsection{Korean Mission (NEXTSAT-1): NISS for the Demonstration of Infrared Spectro-Photometry in Space}

NISS is the near-infrared instrument optimized to NEXTSAT-1, which provides a unique spectro-photometric capability in the near-infrared range. The scientific objectives are to study the cosmic star formation history in local and distant universe through observations of a variety of diffuse objects from star-forming regions in our Galaxy to the light from distant galaxies. The wide spectral range from 0.95 to 2.5 $\mu$m and wide field of view of 2$^\circ\times$2$^\circ$ are to maximize the scientific outputs. Due to the limited pointing information, the low spectral resolution of 20 is designed to study the strong infrared spectral lines related to star formation of our Galaxy, e.g., Paschen-$\alpha$, H$_{2}$ 1-0 S(1) and to measure the large-scale fluctuation of cosmic infrared background radiation in the low-background regions. 

Especially, unveiled near-infrared extragalactic background light can be addressed from the spectro-photometric observations of NISS. The extragalactic background light (EBL) is caused by the light emitted from unresolved galaxies in the line of sight, which enables us to trace the history of the universe. The fluctuation measurements of EBL in wide bands have been made through the previous infrared missions \cite{Seo15, Kim19}, which suggested that the spatial scale of the fluctuation peak is around 1 degree. Thus, the spectral measurement of fluctuation from EBL with the NISS observations may reveal the history of the universe associated with galaxy formation from galaxies observed today to the possible galaxy population contributing to EBL. 

\begin{figure}
	\includegraphics[width=8.0cm]{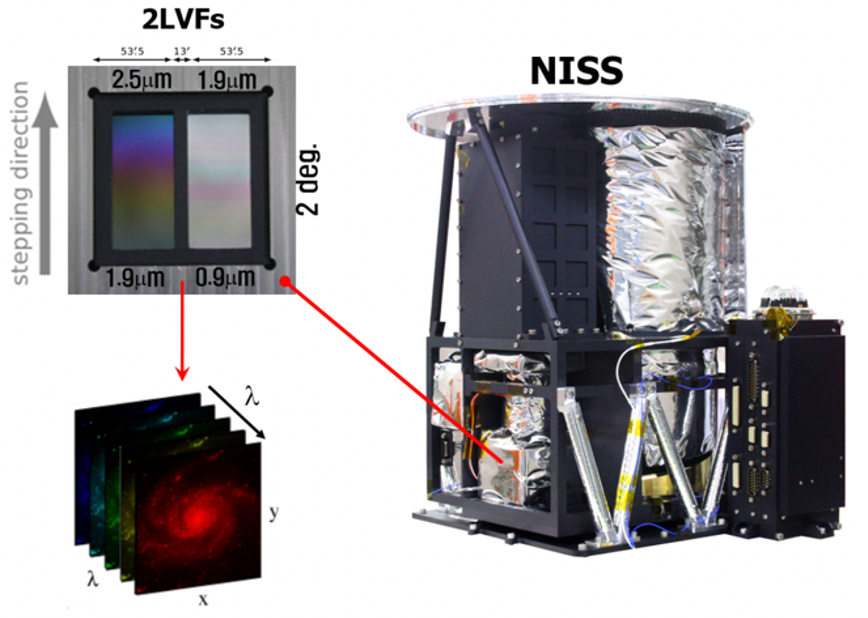}
	\caption{Development of NISS (right),  the operation of LVF (upper left), and the expected output image from NISS (lower left).
	}\label{fig:NISS}
\end{figure}

\begin{figure}
	\includegraphics[width=8.5cm]{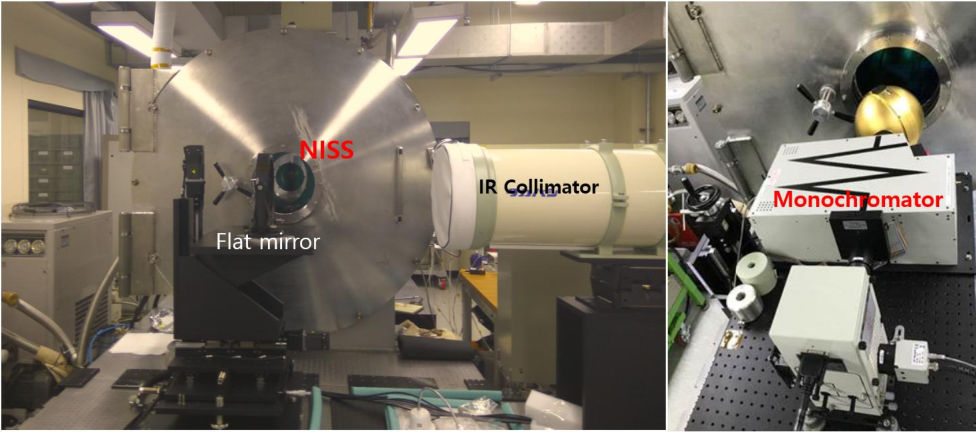}
	\caption{The calibration facility for NISS. The collimator system (left) and the monochromator system (right) are for the purpose of the focus adjustment and wavelength calibration, respectively.
	}\label{fig:NISS-Calib}
\end{figure}

Fig. \ref{fig:NISS} shows the configuration and the expected three dimensional data set of NISS composed of two spatial and one spectral dimension. The central region of $13\rm'$ is blocked to calibrate the dark level in observations. Careful attention was paid to the design of NISS to in order to reduce the volume and the power consumption and to maximize a total throughput. The off-axis optics was newly implemented to cover a wide field of view and a wide wavelength range. The combinations of filters (LVFs and two cut-off filters) try to suppress a leakage of transmission below 0.01\% in out-of-band range. The mechanical structure was tested under the launching and operating conditions for passive cooling of the telescope at 200K stage and stable opto-mechanical support of two off-axis mirrors and relay-lens. The dewar to operate the infrared detector and filters at 80K stage was fabricated and tested as compact as possible. The calibration process in the laboratory is important to measure the performance as well as to obtain the calibration data of the instrument. As seen in Fig. \ref{fig:NISS-Calib}, KASI had constructed the test facilities for NISS, e.g., cryogenic chamber for optical test under the space environments, monochromator system for a wavelength calibration and infrared collimator system for a focus adjustment.

NISS was successfully launched in December 2018 and tested during the operational phase. Due to the shortened cooler life, the science observations were terminated $\sim$6 months after the launch earlier than expected. However, the capability of spectro-photometry was demonstrated in space and minimum data sets for scientific research were secured (see the example in the top-left of Fig. \ref{fig:NISS-SPHEREx} from the observation of NISS). These pioneering technical demonstrations will help to develop the future infrared space mission.

\subsection{International All-Sky Infrared Spectro-Photometric Mission, SPHERE$\textsc{x}$}

\begin{figure}
	\includegraphics[width=8.5cm]{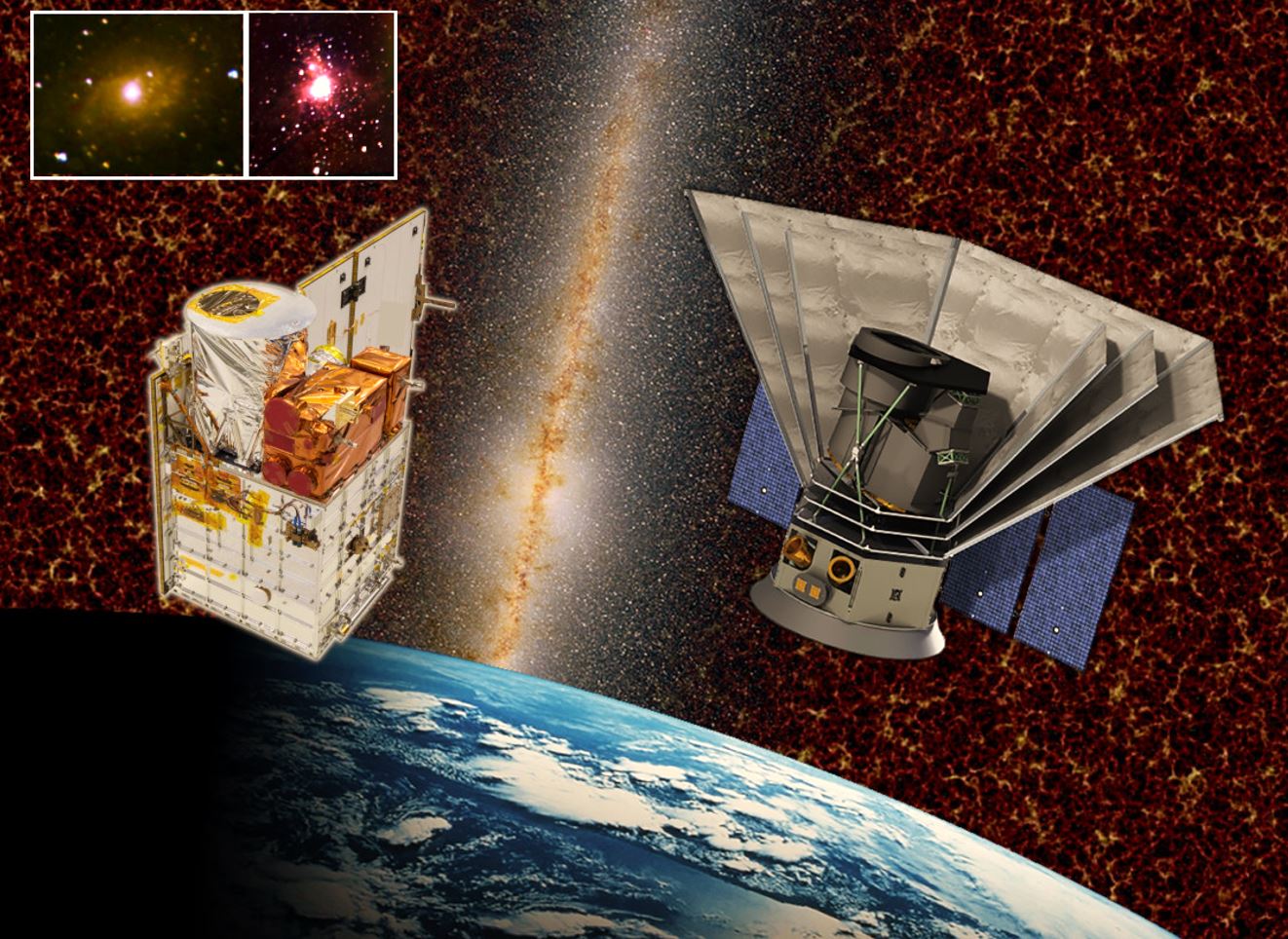}
	\caption{Illustrated image expected from SPHEREx together with NISS onboard NEXTSAT-1 and SPHEREx 
	(modified from the original image from NASA press release on 
	February 14, 2019, {https://www.nasa.gov/press-release/nasa-selects-new-mission-to-explore-origins-of-universe}). 
	The images of Triangulum Galaxy and Large Magellanic Cloud in the top-left are taken from NISS in the spectro-photometric observation.}
	\label{fig:NISS-SPHEREx}
\end{figure}

As an extension of NISS, KASI has been collaborating for the all-sky spectro-photometric survey. The SPEHREx mission (PI Institute: Caltech) will perform the the first all-sky infrared spectro-photometric survey to probe the origin of our Universe, to explore the origin and evolution of galaxies in the two SPHEREx deep fields near ecliptic poles, and to explore whether planets around other stars could harbor life \cite{Olivier16, Olivier18} (see also Fig. \ref{fig:NISS-SPHEREx}). It will also provide the rich legacy archive of all-sky spectral catalog to be used in various research fields, which gives strong scientific synergies with other space missions and ground-based observatories. 

As the only international partner, KASI contributes to SPHEREx in the system calibration as a hardware development and the scientific research. Since SPHEREx implemented an off-axis optics and LVFs for the infrared spectro-photometric survey like NISS, the successful demonstration of NISS in space gives a great benefit in our actual contribution to SPHEREx. Compared to NISS, SPHEREx is designed to have much wider field of view of 3.5$^\circ\times$11.3$^\circ$ as well as wider spectral range from 0.75 to 5.0 $\mu$m. As described in the previous section, KASI already has experience to construct test facilities for the infrared spectro-photometry to make a wavelength calibration and a focus adjustment (see Fig. \ref{fig:NISS-Calib}). These facilities are revised for SPHEREx to test in the lower operating temperature range of below 80K and to calibrate in the wider spectral range up to 5 $\mu$m. 

SPHEREx will be launched in 2023 and it will carry out the all-sky spectro-photometric survey for about 2 years. KASI's participation in the SPHEREx mission is expected to have a great impact on the development of a large space telescope in the future as well as the astrophysical research in the Korean astronomical community.

\section{Toward Next Generation Coronagraph}\label{NGC_sec}

The Sun is extremely dynamic and continuously influences the Earth via the interplanetary magnetic field and solar wind flow.
A sudden release of a tremendous amount of energy from the Sun causes disturbances in Earth's magnetosphere and atmosphere, impacting the performance of modern technologies. For example, it has been known that energetic coronal mass ejections (CMEs) can cause disruption to communications, malfunctions of a navigation system, and radiation hazards to humans in space and airplanes flying in the polar route.
So far, previously flown coronagraphs (e.g., Large Angle and Spectrometric Coronagraph: LASCO) measured only the density of electrons in the solar corona. They have contributed to increasing our knowledge about the CME structures (e.g., Illing $\&$ Hundhausen \cite{1985JGR....90..275I}), dynamics (e.g., Yashiro et al. \cite{2004JGRA..109.7105Y}), their geo-effectiveness (e.g., Kim et al.\cite{2008ApJ...677.1378K}).

The solar wind plays a significant role in the propagation of CMEs that propagate through its quasi-steady flow.
Over the years, the vast CME propagation models for space weather forecasting have relied on incomplete information about the status of background solar wind flows. Due to a lack of observational constraints, the models use semi-empirical techniques (e.g., Wang $\&$ Sheeley \cite{1990ApJ...355..726W}) and global-scale 3D numerical simulations of the corona and solar wind \cite{2001ApJ...546..542L}. 
It has been known that fast solar wind originates in coronal holes, and slow solar wind may be associated with coronal streamers. However, the answers are not yet known about the origin of solar wind on the Sun and how the nascent solar wind is accelerated to several hundred kilometers per second within 10 solar radii (Rs). Precise properties of the sources of the solar winds and their acceleration mechanisms are not known yet.

KASI has been developing a next-generation coronagraph (NGC) in collaboration with NASA. The joint mission is designed to deliver a comprehensive data set to answer the questions, including: Are there signatures of hot plasma released into the solar wind from previously closed fields? What are the velocities and temperatures of the density structures that are observed so ubiquitously within streamers and coronal holes? The new coronagraph images the photospheric light Thomson-scattered by electrons with different density, temperature and speed in solar corona. The observation is based on the study by Cram \cite{1976SoPh...48....3C}
who proposed that the temperature determines the shape of the coronal spectrum, and the bulk flow speed causes the spectrum redshifts. The electron temperature can be measured from the intensity ratio at the temperature-sensitive wavelengths, and the speed can be determined from the ratio of nodal (temperature-insensitive) intensities. Such measurements using the passband ratio imaging have been conducted during total solar eclipses \cite{1996PASJ...48..545I,2003ApJ...599..596R,2018SoPh..293...82R,2020JKAS...53...87C}.

Unlike traditional coronagraph, NGC will obtain density, temperature, and speed of coronal electrons simultaneously beyond the heights where the signal-to-noise ratio to obtain the parameters is not enough from total solar eclipse. There are no routine measurements of all three physical properties in the coronal height between 3 and 8 Rs, where the CMEs are triggered, and the solar wind is accelerated. Given its capability, the new coronagraph is interesting and expected to give critical insights into the fundamental and practically important questions in solar physics. A more detailed description of the scientific context and the measurement technique can be found in Cho et al.~\cite{2017JKAS...50..139C}, Goplaswamy et al.~\cite{10.1117...12.2323807}, and references therein.

A project plan for the development of the NGC had been discussed between scientists from KASI and NASA/Goddard Space Flight Center (GSFC) since 2013. The project got momentum after the formation of the KASI-NASA working group for Heliophysics since 2016.
The working group agreed on the timelines and set responsibilities of both institutes for NGC development and flying it on ISS.

\begin{figure*}[t]
\includegraphics[width=12cm]{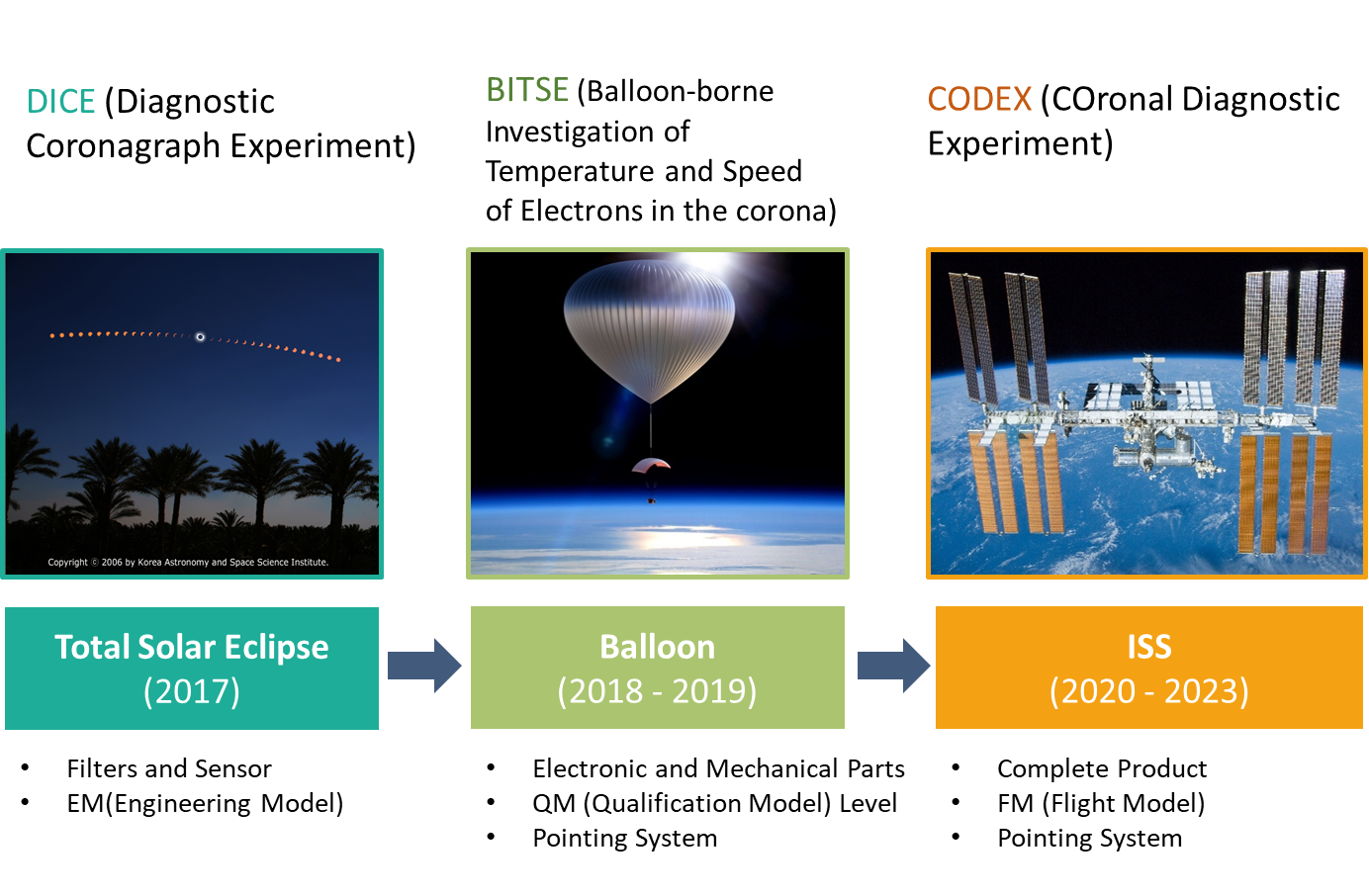}
\caption{(Color online)  Timeline toward the next generation white-light coronagraph.
Observations of total solar eclipse (DICE) and balloon experiments (BITSE) were preformed for the demonstration and validation of the
coronagraph technology before launch of the ISS coronagraph (CODEX).
}\label{CODEX_fig1}
\end{figure*}

As shown in Fig. \ref{CODEX_fig1}, the working group planned three steps: a demonstration of the technology in 2017 (DICE: DIagnostic Coronagraph Experiment), increasing the technology readiness level in 2019 (BITSE: Balloon-borne Investigation of the Temperature and Speed of Electrons in the Corona), and development of the coronagraph (CODEX) to be flown on ISS in 2023. The BITSE and CODEX are NASA-KASI joint missions funded by KASI and NASA.
Table \ref{CODEX_table1} summarizes the mission requirement of CODEX. In this overall project, NASA and KASI will design, build, and perform quality testing of the coronagraph together. Specifically, NASA designs, builds, and qualifies the optical and mechanical assemblies of the coronagraph, and provides engineering facilities and services as needed. KASI develops filter wheel, filters, main electronics, camera, and ground and flight software.

\begin{table}
\caption{Mission Requirement}
\begin{ruledtabular}
\begin{tabular}{llc}
Measurement & Requirement \\
\colrule
FOV (Rs)  & 3.0 - 8 Rs \\
Observable & Polarized brightness (pB) images of corona\\
Filters & 393.5, 405.0 nm (temperature) \\
        & 398.7, 423.4 nm (speed)\\
Cadence & 48-min sequences (temperature $\&$ speed) \\
        & 1-min (density) \\
Spatial Resolution & 5'$\times$5' (temperature) \\
        & 5'$\times$20' (radial direction for speed) \\
        & 29.6" (superpixels) \\
Sensitivity & 0.5 - 2 MK (temperature) \\
        & 0 - 700 km s$^{-1}$ (speed) \\
Life time & 6 month (threshold), 2 years baseline \\
\end{tabular}
\end{ruledtabular}
\label{CODEX_table1}
\end{table}

\subsection{DICE}

\begin{figure}
\includegraphics[width=8.0cm]{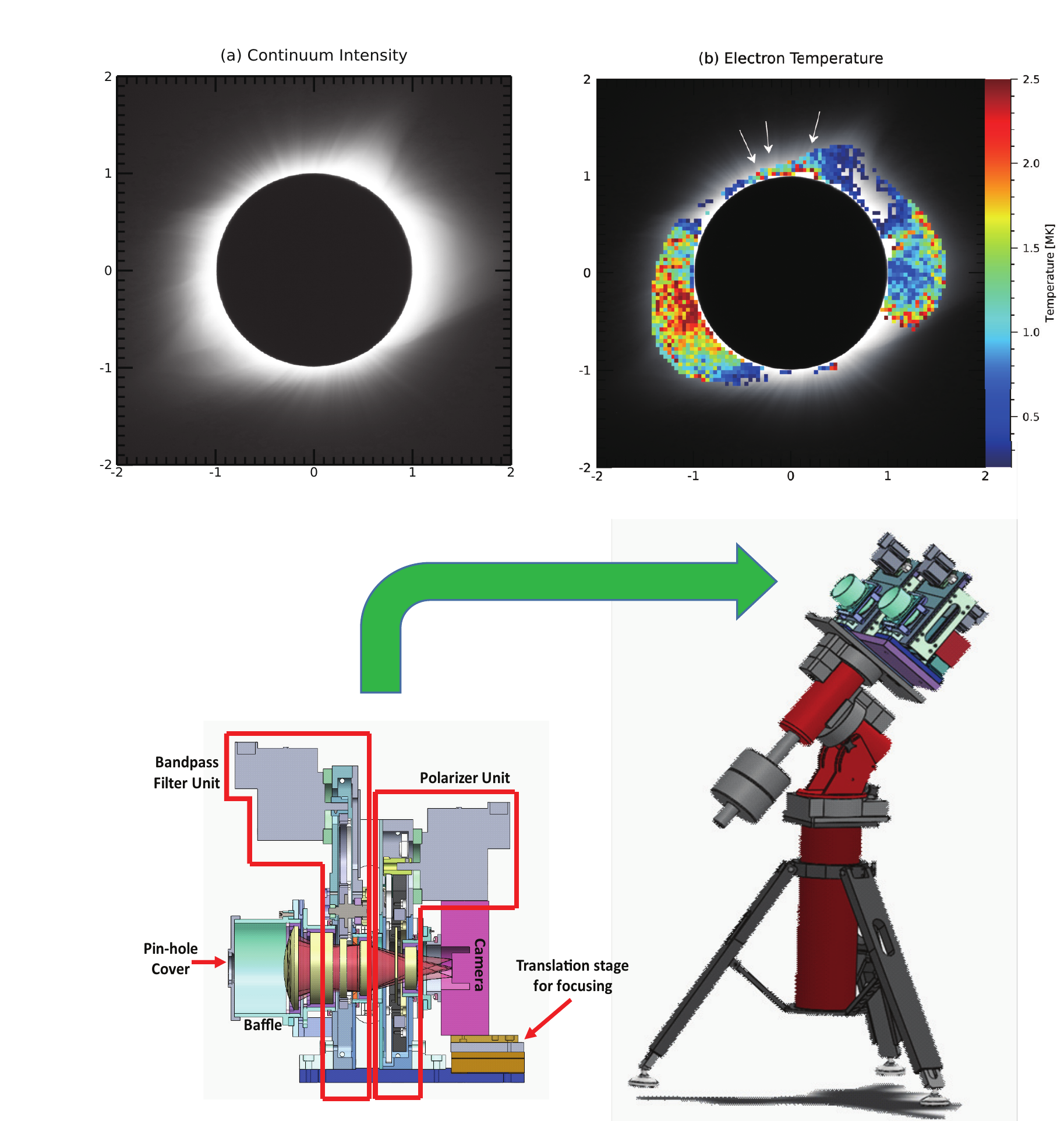}
\caption{(Color online) (Upper left and right panels) Coronal continuum intensity and electron temperature map.
(Lower panel) layout of DICE mounted on the tracking system \cite{2020JKAS...53...87C}. 
}\label{CODEX_fig2}
\end{figure}

KASI coronagraph team built the eclipse observation system, DICE, composed of two identical optical systems installed on a single mount, as shown in Fig. \ref{CODEX_fig2}. The team organized an expedition for the total solar eclipse of August 21, 2017, and observed the solar coronal using the system at a site near Jackson Hole, Wyoming. It was conducted in the limited total eclipse time of about 140 seconds, and polarization data at four different wavelengths were successfully obtained. Due to strong emission from prominences located in the western limb, it could not get information on the coronal electron temperature at the streamer on the west limb. However, the team determined temperature distributions in the eastern limb and northern polar region, and found that temperatures are higher at the streamer and lower at the boundaries between the streamer and the coronal holes, as shown in Fig. \ref{CODEX_fig2}. Since hot spots in the northern coronal hole were identified, the team is investigating the spots in detail to study their relationships with jets and flows in polar plumes. Through the expedition using DICE, the KASI team successfully demonstrated the measurement technology and learned that NGC should be designed and tested carefully to achieve its scientific purpose. More detailed descriptions of the DICE instrument and observation can be found in Cho et al. \cite{2020JKAS...53...87C}. 

\subsection{BITSE}

KASI and NASA coronagraph team had collaborated for a high-altitude scientific balloon experiment.
Two minutes of observation of the total solar eclipse was extended more than 4 hours through new technology demonstration experiment, BITSE.
It was launched on 2019 September 18 (08:50 AM, MDT) from Fort Sumner, New Mexico station of Columbia Scientific Balloon Facility (CSBF).
It reached $\sim$ 40 km in altitude after $\sim$ 2.5 hours flight by the 39 million cubic feet balloon provided by GSFC/WFF Balloon Program Office. BITSE observed the Sun for about five hours and landed at a location near Farmington, New Mexico.
The joint team obtained about 4200 images (60 sets of 70 exposures) in each bandpass filter.

\begin{figure}
\includegraphics[width=8.0cm]{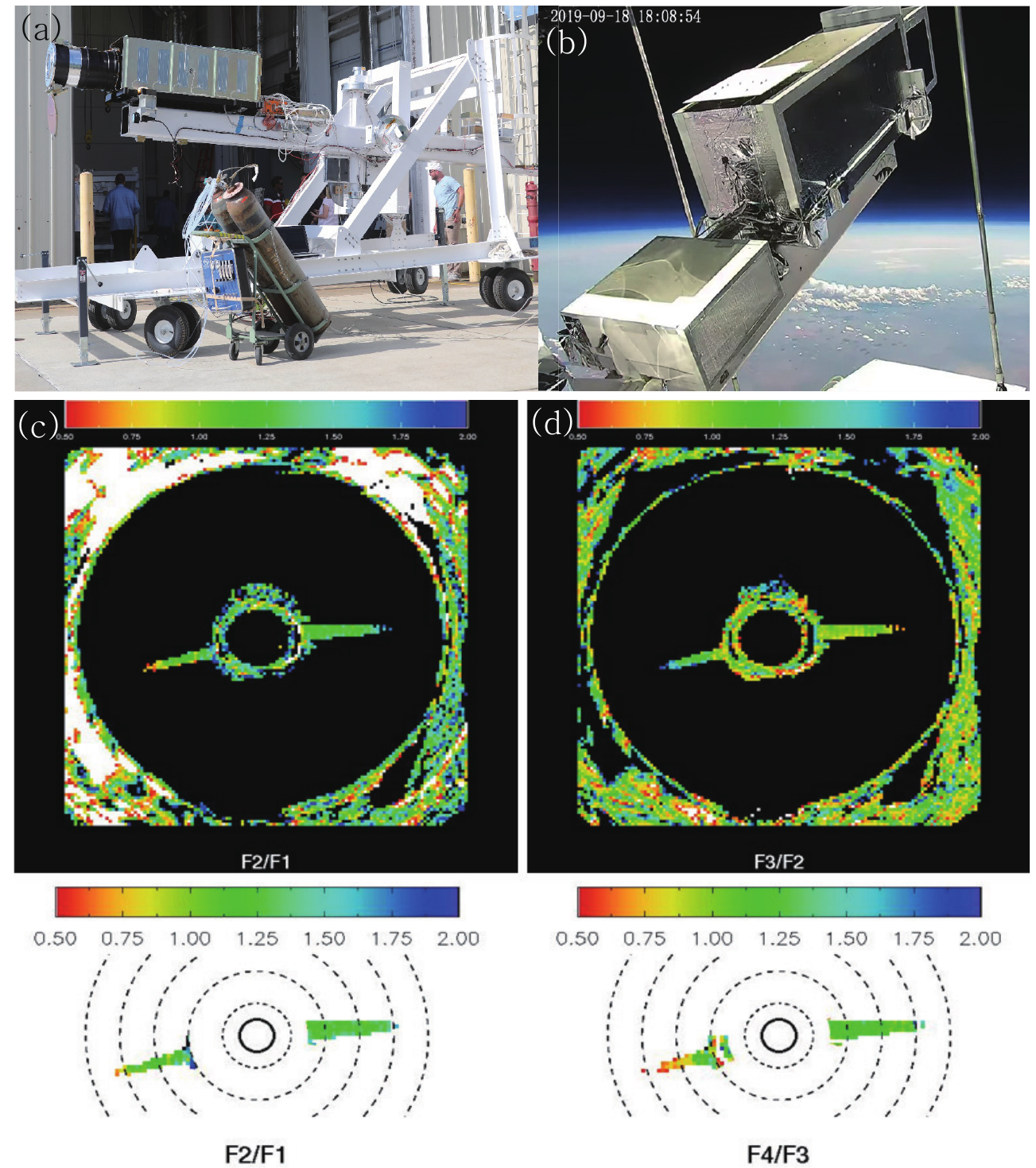}
\caption{(Color online) (a) BITSE system for a Sun-pointing test. (b) BITSE pointing to the Sun at 18:48 UT at the float altitude.
Filter ratio map for (c) electron temperature (4050~\AA / 3935~\AA) and (d) speed (4234~\AA / 3987~\AA).
Regions outside the streamers were masked out due to no signal there \cite{BITSE}. 
}\label{CODEX_fig3}
\end{figure}

\begin{figure*}[t]
\includegraphics[width=12.0cm]{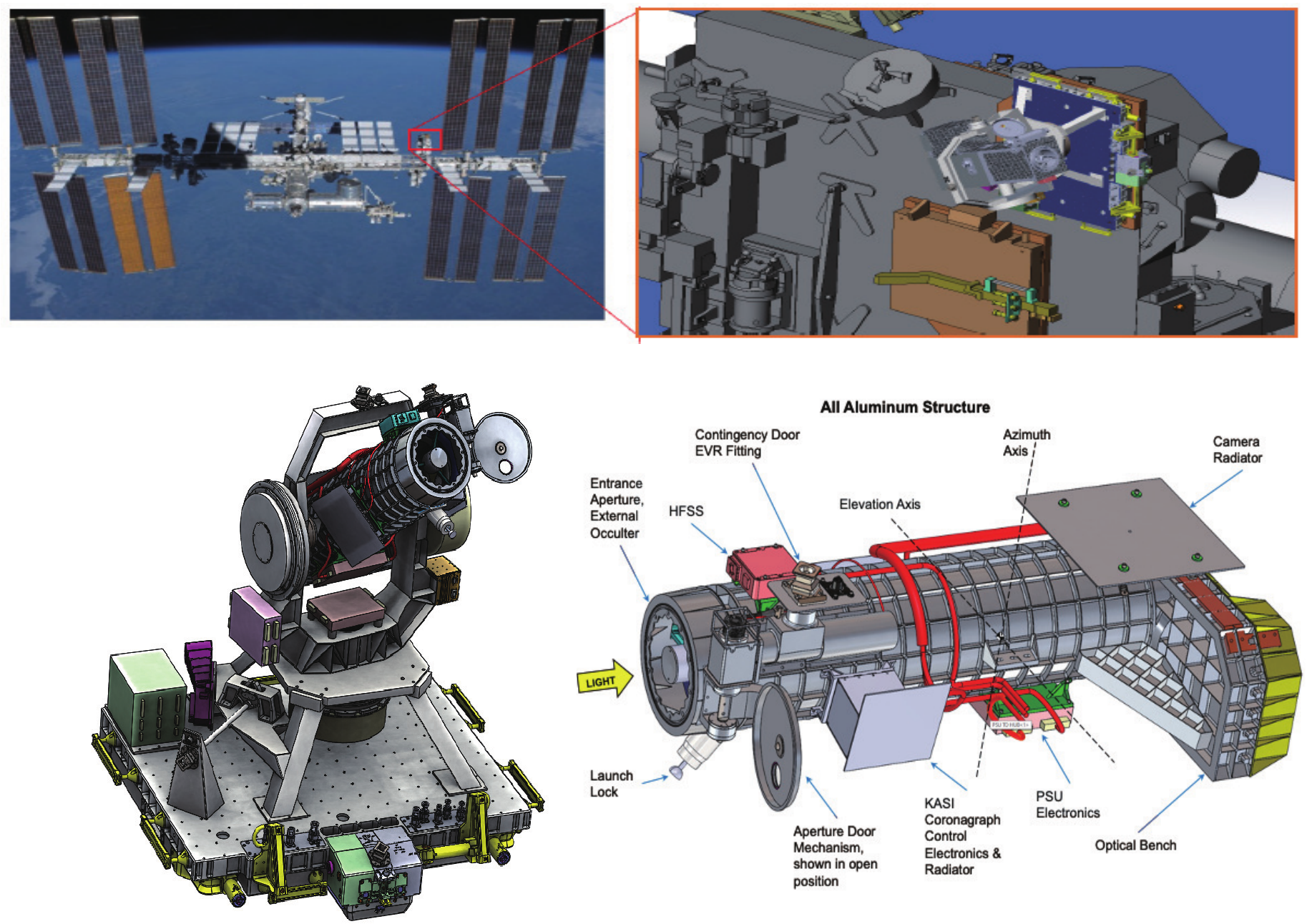}
\caption{(Color online) (Upper panels) Installation site (ELC3-3) of CODEX.
(Lower panels) Payload layout and coronagraph configuration.
}\label{CODEX_fig4}
\end{figure*}

Fig. \ref{CODEX_fig3} shows BITSE integrated to the gondola and the Wallops Arc Second Pointing (WASP) system at NASA's Wallops Flight Facility (WFF).
BITSE observed the Sun from 18:48 UT to 22:11 UT at the altitude of $\sim$ 39 km over a period of $\sim$ 4 hrs.
As shown in Fig. \ref{CODEX_fig3} (c) and (d), BITSE obtained filter ratios of temperature and speed for two streamers on the east and west side of the Sun.
For the main features, the BITSE team determined temperature and speed values using the filter ratios,
and they are working on the atmospheric attenuation at BITSE wavelengths to confirm the result.

BITSE is the first coronagraph observing from a stratospheric balloon in the blue end of the solar spectrum.
Through the mission, two critical technologies were demonstrated.
The first one is narrowband filters that block all wavelengths of visible light except for those in four spectral bands in the violet range.
The second one is the polarization camera that can directly collect polarized light without an extra mechanism for rotating polarization wheels as done by the traditional detector.

BITSE demonstrated that it could get images of the solar corona and obtain spectral polarization brightness images needed to determine the temperature, speed, and density in the corona. The NASA-KASI coronagraph team achieved TRL 6 or higher in all technologies for NGC through the BITSE mission. More detailed description on the mission and preliminary results can be found in Goplaswamy et al. \cite{BITSE}.

\begin{figure*}[t]
\includegraphics[width=12.0cm]{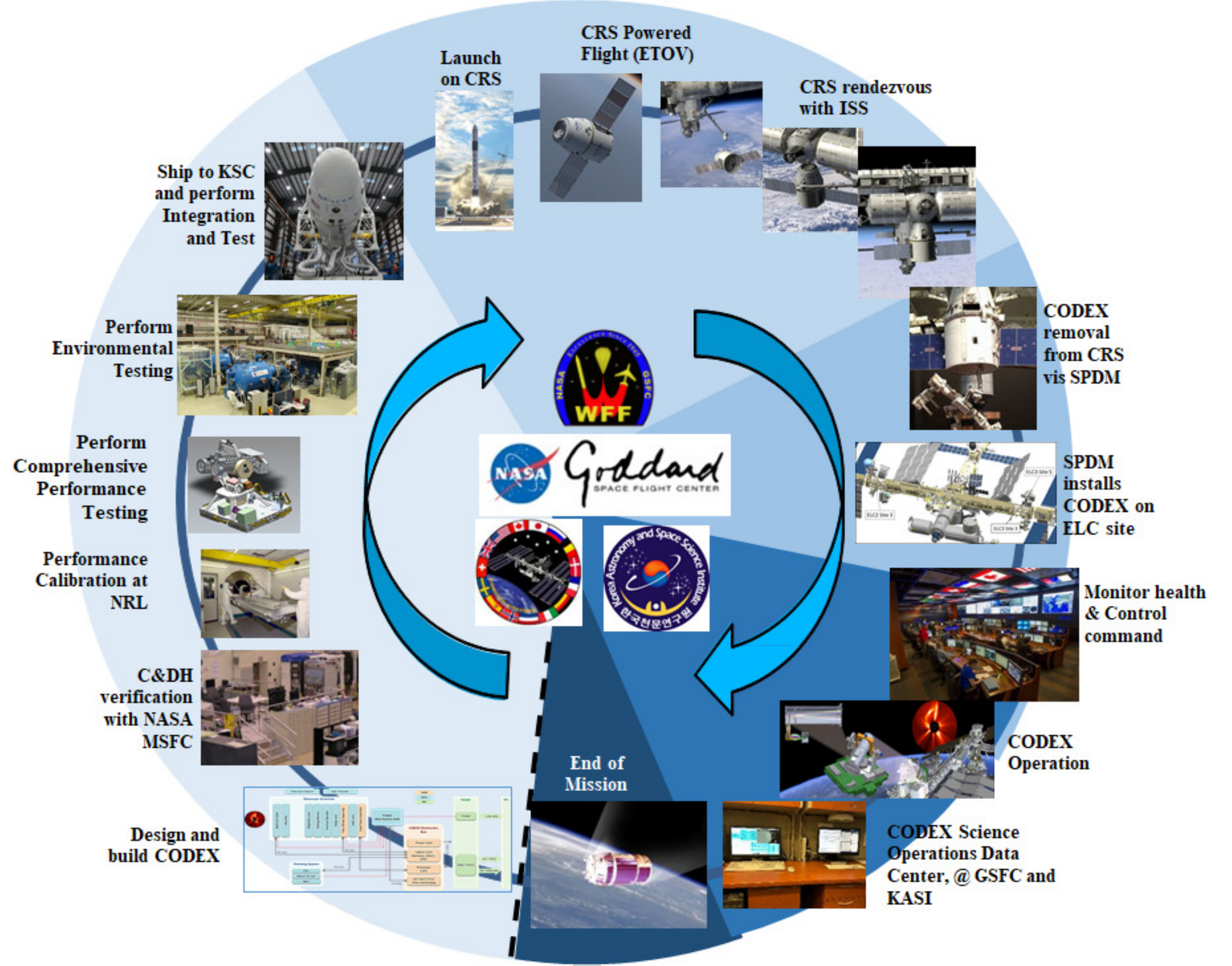}
\caption{(Color online) Mission phases and key activities that follow standard mission development and ISS practice.
}\label{CODEX_fig5}
\end{figure*}

\subsection{CODEX}

The BITSE flight duration was too short, and the background sky was still too bright to obtain all the data needed for achieving the science objectives. The next step to utilize the new coronagraph technology demonstrated by BITSE is to fly in space. The KASI-NASA coronagraph team is developing the next generation coronagraph, CODEX utilizing the ISS platform and Azimuth/Elevation (Az/El) Telescope Gimbal Mount.

As shown in Fig. \ref{CODEX_fig4}, CODEX utilizes existing ISS infrastructure and standardized mission designs for visiting payload.
ISS is a cost-effective platform for the new coronagraph since it provides ample power and a telemetric margin.
Main parts of the coronagraph are aperture door, external occulter, straylight baffles, heat rejection mirror, filter wheel, polarization camera, and control electronics.
CODEX system consists of the coronagraph part and other components, i.e., pointing, power, command and data handling (C$\&$DH), and thermal system.
It uses passive cooling (radiator) and cold bias systems and heaters for fine temperature control.

Fig. \ref{CODEX_fig5} presents mission phases and critical activities of CODEX from pre-launch to the end of the mission.
It is expected to be launched in February 2023 in Commercial Resupply Services (CRS) vehicle and installed onto zenith-pointing location (ELC-3-3) on ISS as shown in Fig. \ref{CODEX_fig4}. The expected orbit altitude is about 400 km, and the orbit period at 400 km is 92 mins. Orbit inclination is $\sim$ 51.6 degrees, which allows for excellent solar viewing. CODEX shall produce more than 1000 density images with less than 1-minute cadence per day, 16 temperature images per day from the average of less than one hour observation of each orbit, and two sets of three-hour average speed images per day. For higher signal-to-noise ratio, images may be further averaged over more orbits. It is a six-month minimum mission with a two-year baseline mission option to begin solar cycle evolution and variability studies.

It is expected that CODEX will conduct simultaneous measurements of electron density, temperature, and speed.
The new data from CODEX will stimulate advances in scientific modeling capabilities to improve modeling of the heliosphere.

\section{SNIPE}\label{SNIPE_sec}

\begin{table*}[t]
\caption{SNIPE Science Payload}
\begin{ruledtabular}
\begin{tabular}{cccc}
Instruments & Range & Cadence & Heritage \\
\colrule
 \multirow{2}{*}{Langmuir Probe} & \multirow{2}{*}{$2\times10^3 - 2\times10^6 cm^{-3}$} & 0.1s for Ne & \multirow{2}{*}{STSAT-1\cite{2012AnGeo..30.1567L}}  \\
 &&1 s for Te &\\
 \\
 \multirow{2}{*}{Magnetometer} & $\pm$60,000 nT &  \multirow{2}{*}{0.1 s} &  \multirow{2}{*}{STSAT-1\cite{2012AnGeo..30.1567L}} \\
 &sensitivity: 30 pT/$\sqrt{Hz}$ at 1Hz &&\\
 \\
  \multirow{2}{*}{Solid State Telescope} &  100 - 400 keV (Variable) electron & 0.01s for Burst &  \multirow{2}{*}{STSAT-1\cite{2012AnGeo..30.1567L}}\\
  &16 energy channels & 0.1 s for Survey & \\
\end{tabular}
\end{ruledtabular}
\label{SNIPE_table1}
\end{table*}



\begin{figure*}[t]
\includegraphics[width=16.0cm]{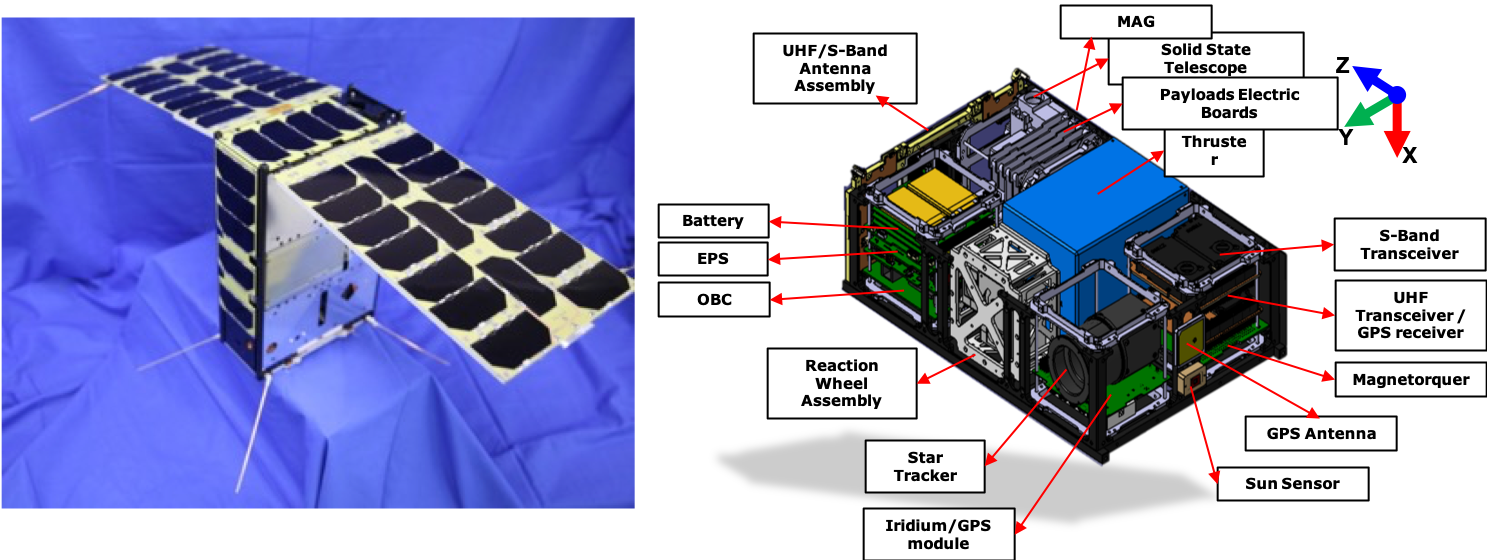}
\caption{The photo of the SNIPE EQM (left), and the configuration of spacecraft subsystems (right).
}\label{SNIPE_fig1}
\end{figure*}

SNIPE is a scientific mission consisting of four 6U CubeSats of $\sim$10 kg under development by KASI for space weather research. The observation of particles and waves causing space storms with a single satellite inherently suffers from space-time ambiguity. Recently, such ambiguity has often been resolved by multi-satellite observations such as CLUSTER\cite{2001AnGeo..19.1197E}, THEMIS\cite{2008SSRv..141....5A}, MMS \cite{2016SSRv..199....5B}, and SWARM\cite{2019GeoRL..4612731L} missions; however, the inter-satellite distances of these satellites were generally vast ($>$ 100 km). Hence, small scale plasma phenomena on low altitude orbit have not been studied intensively. The goal of the SNIPE mission is to measure the spatial and temporal variations of the micro-scale plasma structures on the topside ionosphere. For example, polar cap patches, field-aligned currents (FAC), radiation belt microbursts, and equatorial and mid-latitude plasma blobs and bubbles are the main observation target of the SNIPE mission. Measurements of the dimensions, occurrence rates, amplitudes, and spatiotemporal evolution of such small-scale plasma structures will determine their significance to the solar wind-magnetosphere-ionosphere interaction and quantify their impact on space weather. 

The SNIPE’s 6U CubeSats have been developed with the cooperation of KASI and KARI. The spacecraft has two deployable solar panels and four body-mounted panels to produce electric power more than 44W. SNIPE is a 3-axis stabilized spacecraft consisting of three reaction wheels, magnetorquers, Sun sensors, star trackers, MEMS Gyros, and magnetometers to achieve the capability of altitude control of the accuracy of less than 1 arc-degree. The on-board computer utilizes the core Flight Software (cFS) and handles the command and data in addition to the guidance, navigation, and control processing with a high-speed micro dual processor. The communication subsystem is a critical part of deciding CubeSat’s success. For the low-speed but effective communication, UHF, and for the high-speed data download ($\sim 2$ Mbps), S-band uplink/downlink RF modules are installed on the bus system. In addition, the IRIDIUM communication module would make possible the real-time housekeeping monitoring in addition to providing an opportunity to upload changes in operational modes when geomagnetic storms occur. In the SNIPE mission, the key technology is the demonstration of formation flying. For the orbit maneuver, each spacecraft has a cold gas thruster that produces the total thrust of del-v 50 m/sec, and GPS receivers to determine the accurate position of less than 10 m. Fig. \ref{SNIPE_fig1} shows the picture of the SNIPE Engineering and Qualification Model (EQM) and the configuration of spacecraft subsystems.

\begin{figure}
\includegraphics[width=8.0cm]{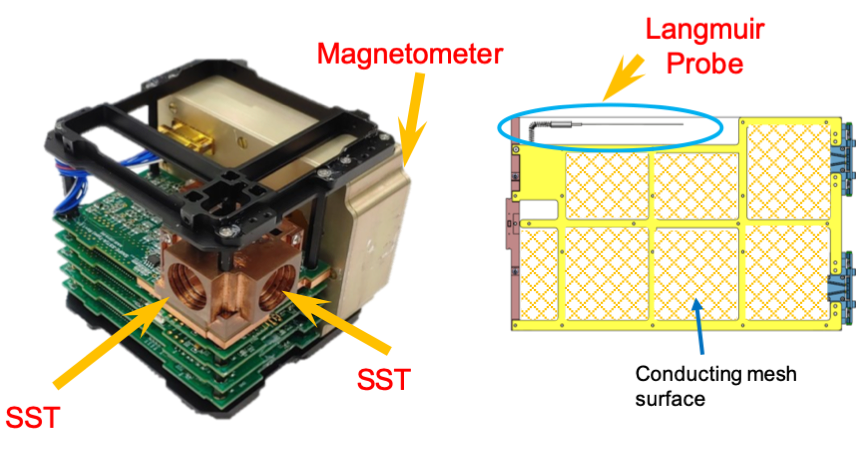}
\caption{Images of scientific payloads, Solid State Telescope, Magnetometer, and Langmuir Probe.
}\label{SNIPE_fig2}
\end{figure}

Each SNIPE CubeSat carries the same scientific payloads, a Langmuir probe, a magnetometer, and a solid-state telescope (SST), which are the heritage from STSAT-1. 
Table \ref{SNIPE_table1} lists the instruments and their capabilities, and Fig. \ref{SNIPE_fig2} shows the image of scientific instruments. 
The Langmuir probes measure ionospheric plasma densities and temperatures. When a seeping voltage is applied to the metal rod exposed to the plasma environment, the current is measured. From the current-voltage characteristic curve, the plasma densities and temperatures are derived. The Langmuir probe produces electron density data every 0.1 s, and electron temperature every 1 s. Because the satellites move with supersonic speed, the spacecraft body generates different plasma environments in the backside track (wake). To reduce the effects caused by the wake, the probes are placed on the backside of the solar panel and deployed in orbit. 

The SST consists of two Passivated Implemented Planar Silicon (PIPS) detectors whose active area is 50 ~mm$^2$. The detectors measure the energetic electron flux of 100-400 keV with the 16 energy channels. Two collimators identify the incident direction of particles, one is perpendicular and the other is parallel to the geomagnetic field. In order to observe small scale electron precipitation phenomena on polar region, the SST has fast time resolution, 10 ms for burst mode and 100 ms for survey mode with the dead time of 10 $\mu$sec. The geometric factor of SST is 0.02~cm$^{-2}$~sr$^{-1}$. The flux gate magnetometers are aboard SNIPE to measure the vector geomagnetic field in the range of $-60,000$ to $+60,000$~nT with the 24-bit AD converter in frequency range of DC -- 10 Hz. The magnetic field data are used to identify field-aligned current variation associated with crossings of the auroral oval and transient current loops. To cancel the internal noise caused by electric currents in the bus system, two tri-axial sensors are installed. The noise level of the magnetometer is 30 pT/$\sqrt{Hz}$ at 1 Hz. Compare it with other missions, THEMIS (10 pT /$\sqrt{Hz}$)\cite{2008SSRv..141..235A} and Ex-Alta-1 (100 pT/$\sqrt{Hz}$)\cite{2016JGRA..12111839M}. 

\begin{figure}
\includegraphics[width=8.0cm]{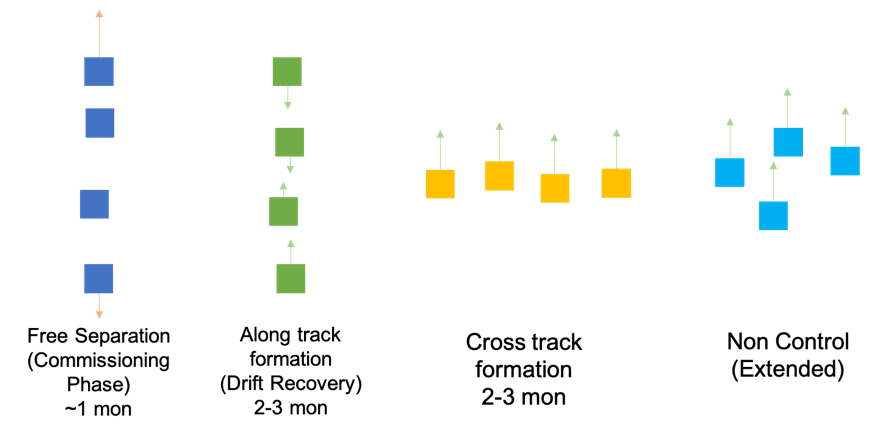}
\caption{The configuration of SNIPE mission formation flying
}\label{SNIPE_fig3}
\end{figure}

Table \ref{SNIPE_table2} lists SNIPE mission phases, and Fig. \ref{SNIPE_fig3} shows the configuration of SNIPE formation flying. During the 5-day Launch and Early Orbit Phase (LEOP), SNIPE will undergo separation, stabilization, Sun acquisition, and on-orbit checkout. The thrusters will be calibrated and the payload checked during the 25-day In-Orbit Testing (IOT) phase. Yonsei University will plan formation flying during the Science Mission Phases (SMP). During SMP-1, the along-track configuration will steadily diminish over three months from an array of spanning ~2000 km to spanning 10 km. During SMP-2, cross-track separation distances will steadily increase over 3 months until the array spans 400 km. During the extended mission, expected to last at least 6 months, no control will be exercised over the evolving orbits. Science operations will continue until the spacecraft cease to operate. Throughout the science and extended phases, one of each 15 orbits will be for downlink, one for thruster operation, one for momentum dumping, eight for Langmuir Probe and magnetometer operations, three for survey mode (0.1s time resolution) SST operation, and 1 for fast mode (0.01s time resolution) SST operation. The 500 km altitude orbital period will be ~96 min, yielding velocities of 7.5 km/s.

\begin{figure*}[t]
\includegraphics[width=12.0cm]{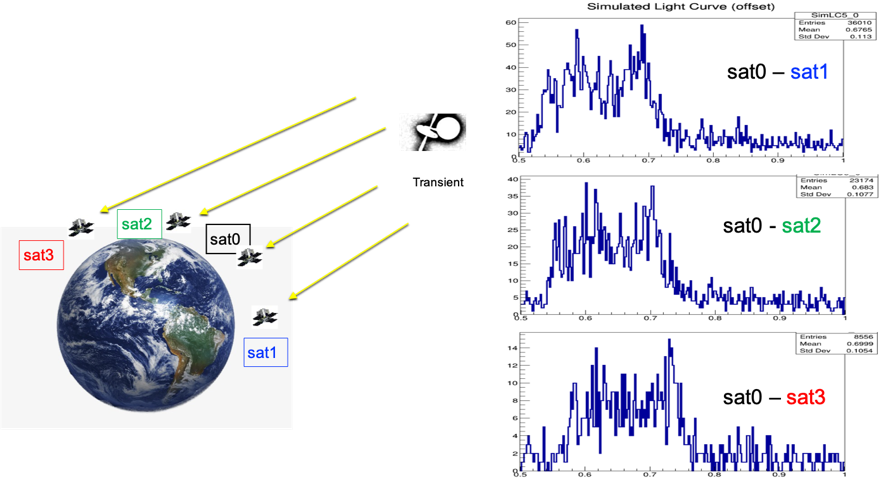}
\caption{Conceptual diagram for localization of gamma-ray bursts. Four CubeSats detect delayed gamma-ray count rate and identify the direction of the gamma-ray burst position.
}\label{SNIPE_fig4}
\end{figure*}


\begin{table*}
\centering
\caption{SNIPE Mission Phases}
\begin{ruledtabular}
\begin{tabular}{lll}
Mission Phase & Duration & Spacing \\
\colrule
LEOP & 5-Day & Free-flying, spans 300 km along track  \\
IOT & 25-day & Free-flying, spans 2000 km along track\\
SMP-1 & 3 months & Inter-spacecraft distances diminish to 10 km \\
SMP-2 & 3 months & Inter-spacecraft distances increase to 400 km cross track \\
\end{tabular}
\end{ruledtabular}
\label{SNIPE_table2}
\end{table*}

Besides the space weather research, the SNIPE mission has another astrophysical objective, detecting GRBs. Weber et al. \cite{2018SPIE10699E..2PW} proposed the Cubesats Applied for MEasuring and LOcalising Transients (CAMELOT) mission that is the constellation of nine 3U CubeSats equipped with CsI scintillator-based gamma-ray detectors. By cross-correlating the light curves of the detected GRBs, the fleet shall be able to determine the time difference of the arriving signal between the satellites and thus determine the position of bright short bursts with an accuracy $\sim 100'$ (Fig. \ref{SNIPE_fig4}). To demonstrate the technology of the CAMELOT mission, CsI gamma-ray detectors combined with GPS and IRIDIUM communication modules are placed on each SNIPE CubeSat. The time of each spacecraft is synchronized, and when the GRB is detected, the light curve will be transferred to the Mission Operation Center (MOC) by the IRIDIUM communication module. By measuring the time difference of each GRB signals, the technology for the localization of GRB will be qualified. If we have good results, we will propose the new astrophysical mission for investigating the origin of GRB.

Work on SNIPE began in January 2017. Phase A (System Design) was complete by October 2017 and Phase B (Critical Design) by November 2019. The mission is currently in Phase C (Integration), nominally awaiting launch from Baikonur on a Russian Soyuz-2 rocket as a secondary payload to the primary Korean Compact Advanced Satellite (CAS500-2) mission. Delays on CAS500-2 indicate that SNIPE will be launched during the third (Q3) or fourth quarter (Q4) of CY 2021.

\section{UVOMPIS Plan}\label{UVOMPIS_sec}


We introduce the UVOMPIS payload for the CAS500-3. The CAS500 is a 500 kg-class Compact Advanced Satellite, to cope with the demand for various space missions by providing a standard bus platform. As a KSLV payload, the CAS500-3 will perform the space astronomy and space science, and demonstrate the space core technologies. The development of bus system of the CAS500-3 has started in 2019, while the payloads will start in 2021.
The UVOMPIS is a unique instrument to observe the universe from ultraviolet to optical band with polarizing capability. It is required not only to research the novel scientific purposes but also to demonstrate the core technologies for the next Korean space telescope. To do that, we have focused on three issues in the proposals for the payload of CAS500.

\begin{figure}
\includegraphics[width=8.8cm]{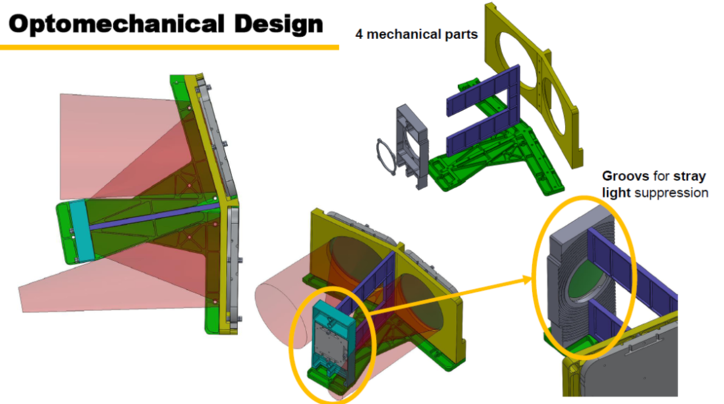}
\caption{
The conceptual design of the UVOMPIS optomechanical part
}\label{UVOMPIS_fig}
\end{figure}

First, the camera should cover wide-range of wavelengths to meet the requirements for the science cases. The UVOMPIS has four bands from 300 nm to 800 nm on two detectors, which is optimal to investigate the SED (Spectral Energy Distribution) of the universe. Particularly, we will put a 4-set polarizing filter on each wave band, which makes it possible to research the interstellar dust and the Zodiacal light.
Second, the UVOMPIS is optimized for the LSB (Low Surface Brightness) targets, by adapting the TMA (Three Mirror Anastigmat) telescope. For this type of telescope, we have designed the aperture size to be 20 cm, the field of view 2 $\times$ 4 degrees, and f-number 2, which will have best performance to observe wide low-brightness diffuse objects.   Fig. \ref{UVOMPIS_fig} shows the conceptual design of the UVOMPIS optomechanical system.
Finally, there are some limitations on policies and resources because the CAS500-3 will be the first payload of KSLV.  Therefore, we will use no cooler, no moving part, no ITAR (International Traffic in Arms Regulations) parts. 

We will also develop the UVOMPIS payload in 3 years with the proven technologies and heritages. The overall envelope shall not exceed 400 $\times$ 350 $\times$ 600 mm, while the mass is less than 50 kg. We expect the power consumption of 20 W in average, and 30 W at the peak.
The major scientific purpose of the UVOMPIS is to observe the LSB targets in our Galaxy and in the nearby galaxies. To investigate the Ultra Diffuse Galaxies (UDG), the galaxy clusters, and the Intracluster Light (ICL), we should understand the foreground noise from our Galaxy. One of the most severe and large foreground source is the cirrus clouds. If we detect the cirrus clouds with multibands polarizing imagers, then we will set a new foundation for researching the LSB sciences.

The other important foreground source is the Zodiacal light. The interplanetary dusts originate from the comets, asteroids, planets and satellites. These various types of dust have different physical properties and their distributions vary according to the position in the solar system. Using the UVOMPIS, we can monitor these dusts from the observation results of the Zodiacal light, which shall be the first wide coverage database of it.  
Observing the aurora with the UVOMPIS is another scientific purpose. There shall be a package of particle measurement instruments on the CAS500-3 also. With the combination of these particle information, the UVOMPIS can provide excellent data for analyzing the aurora light according to the incident particles from the magnetosphere.

\section{Conclusion and Prospects}\label{conclusion_sec}

In this review, we described individual space missions in Korea for astronomy and astrophysics which were performed in the past as well as are ongoing now and planned in the future. We included missions that Korean astronomers/astrophysicists not only led, are leading, and will lead, but also participated in and are participating in through the international collaboration. Most Korea-leading missions in the past were successful in satellite-operating as expected, taking data, and making scientific contributions even if they were done in the beginning phases of Korean space program that faced many challenges. Also in the foreign missions in the past, Korean astronomers/astrophysicists played an important role in not only scientific discoveries but also technological advancements. The experiences obtained from the participation in the foreign missions (as well as from Korea-leading missions themselves) helped Korean space missions to move forward and resulted in the new space mission programs that followed. The ongoing and planned missions became possible based upon the heritage of the previous missions. 

We believe that space missions for astronomy and astrophysics in Korea have a bright future because Korean government continues to promote the space program with the support of Korean astronomy and astrophysics community. However, it seems to us that the future direction of space missions in Korea, particularly those for astronomy and astrophysics, has not been discussed yet either in the policy-decision level of the government or among the astronomy and astrophysics community. We feel that we have now reached the moment to discuss it. As a beginning of this effort, we make some suggestions below based upon our experiences so far. 


For the next space missions, Korea may need to take two-track strategy. First, we need to actively attempt to be involved in international, large space missions. For example, as currently being discussed (and suggested as well at the end of Section \ref{GALEX_sec}), we can pursue to join the Large UV/Optical/IR Surveyor (LUVOIR) mission as a partner. Second, we need to keep developing ideas for our own missions that can be led by Korea but with a relatively limited amount of funds and resources.
This type of space mission should focus on unique ideas that are scientifically beneficial, but can be pursued even with relatively small spacecrafts. At the same time, we also need to develop ideas for a large space mission and for this type of large mission, we continue to persuade the government by appealing the necessity to carry out the large mission led by Korea 
although it must be discussed and agreed first in the Korean astronomy and astrophysics community what missions should be pursued first or with higher priorities. There might be certain limitations for this track mainly due to technological reasons.      
Furthermore, we need to endeavor to secure a stable satellite system that is necessary for a large space mission. In Korea, the space missions for astronomy and astrophysics (and space sciences as well) have been assigned to satellites that are planned for verification of engineering technique, and thus the satellite systems for the science missions have not been necessarily reliable.

As indicated from the list of authors (particularly from their affiliations), a significant fraction of space missions for astronomy and astrophysics have been done and are under operation and planning by KASI. Thus, taking CODEX as an exemplary mission which is a joint-mission between KASI and NASA, we can show how the past and current experiences evolve to the planning in the future as follows.    
%
Through the steps toward the next-generation coronagraph, KASI established a critical international partnership with NASA in heliophysics.
The BITSE KASI and NASA team formed the core of the CODEX team and fulfilled the NASA-KASI working group goals.
Flying CODEX on ISS serves as a pathﬁnder for an improved instrument to ﬂy on the Lunar Orbiter Platform-Gateway or Earth-Sun L4 Space Weather mission as a future mission.
This upcoming platform provides extended solar viewing as an ideal location for performing continued science and long-term space weather requirements.

We also want to draw attention to new windows for future missions in Korea.  
Recent advances in small satellite technology are changing the scale of space missions opening up the possibility of using CubeSat‐type platforms as a solution for low‐cost constellation missions \cite{2018SPIE10699E..2PW}. Particularly in the field of astrophysics and heliophysics, multi-point observation from CubeSat constellation has excellent potential to address many outstanding problems. The experiences obtained in developing the SNIPE mission will pave a new way for space-borne experiments. 
Another new window is to utilize ISS as CODEX does. If Korean community could propose to build (small) detectors/telescopes that are fit for ISS and capable of carrying out unexplored observations, then it would be worth seeking opportunities at ISS. Using ISS has an advantage to avoid the technological issues related to the satellite launching and operation.   

Finally, we may initiate the discussion on the priorities in the future (large) missions in Korea by considering the fact that there have been very few space missions in Korea dedicated to high energy astronomy and astrophysics (i.e., X-ray and gamma-ray) so far. 
Interestingly, however, the first attempt of the space program for astronomy and astrophysics in Korea was through an X-ray rocket mission, the X-ray Detector System (XDR)~\cite{1998JASS...15..373N}. The XDR was developed by KASI for 3 years (1995--1997) and composed of a detector and a signal processing system. The detected data were transferred to the ground using the rocket's telemetry system. On July 9, 1997, XDR was first launched from the Taean rocket flight center onboard the two-stage Korean Sounding Rocket (KSR II) which was developed by KARI. However, unfortunately, data transmission and reception were not made possible due to the rocket's inertial navigation failure. After that, on June 11, 1998, the second KSR II was launched, and it was confirmed that XDR had operated normally for about 1 minute. Although the astronomical data were not obtained, the mission was of great significance in that it was the first domestic space mission for astronomy and astrophysics, in particular, for the X-ray astronomy. It is also worth mentioning that two missions described in this review, UBAT in UFFO-Pathfinder and CSI gamma-ray detectors in SNIPE, are also related to high energy astronomy and astrophysics. 
In spite of this early attempt, the number of astronomers and astrophysicists working in the field of high energy astronomy and astrophysics (particularly in the field of observation) is still very small. The reason for this can be found in the history of Korean astronomy and astrophysics which have been influenced a lot by the Korean economy. 
Historically, Korean astronomy and astrophysics community was established primarily with optical and radio astronomy simply because Korea could not afford to have space missions which are essential for high energy astronomy and astrophysics (and for some other bands). 
%
%
The situations for (far) UV and (far) infrared, which also require space missions, were not very different before the space missions on them began in Korea as described in this review. Thus, we may ask the following question. Is it worth pursuing space missions on high energy astronomy and astrophysics now (or again) with higher priorities than the other missions that are likely to inherit the previous missions? We are not trying to answer this question here, but we hope that we will be able to begin the priority discussion on the future missions by asking this kind of question.  

\begin{acknowledgments}
We would like to congratulate the Astrophysics Division of Korean Physical Society (KPS) on its 25th anniversary and to give a special thanks that this review is included in the special issue.   
%
FIMS/SPEAR was funded by the Ministry of Science and Technology (Korea) and NASA (USA). The FIMS/SPEAR mission was a joint project of KASI, KAIST, and the University of California at Berkeley (USA).
%
MIRIS was support from the National Research Foundation (NRF) grant funded by the Ministry of Science and ICT (MSIT) of Korea (2014M1A3A3A02034746).
YWL acknowledges the support from NRF of Korea (2017R1A2B3002919).  
GALEX was a NASA Small Explorer, launched in April 2003. We gratefully acknowledge NASA’s support
for construction, operation, and science analysis for the GALEX mission, developed in cooperation with the
Centre National d’Etudes Spatiales of France and the Korean Ministry of Science and Technology.
The Korean participation of AKARI was supported by the Phase 1 BK21 program, KOSEF-JSPS Cooporative Program,  and NRF grants (2006-0093852, 2012R1A4A1028713).
MBK and IHP acknowledge the support from NRF of Korea (2018R1A2A1A05022685, 2017K1A4A3015188).
WSJ acknowledges the support from NRF grant funded by MSIT of Korea (2018M1A3A3A02065645).
NGC mission has been funded by the Ministry of Science and Technology (Korea) and NASA (USA). 
The BITSE and CODEX projects were a joint project of KASI and NASA (USA). 
KK acknowledges the support from NRF (2016R1A5A1013277).

\end{acknowledgments}

\bibliographystyle{aip}
\bibliography{bibliofile}

\end{document}